\documentclass[referee]{aastex}
\begin{document}

\title{Pecular velocities of galaxies in the Leo Spur\thanks{ 
Based on observations made with the NASA/ESO
Hubble Space Telescope, obtained at the Space Telescope Science Institute, 
witch operated by the Association of Universities for 
Research in Astronomy, Inc., under NASA contract NAS 5--26555. These 
observations are associated with program SNAP 13442.}}

\author{Igor D. Karachentsev}
\affil{Special Astrophysical Observatory, the Russian Academy of Sciences,Nizhnij Arkhyz,    Karachai-Cherkessian Republic,
   Russia 369167}
   \email{ikar@sao.ru}

\author{R. Brent Tully}
\affil{Institute for Astronomy, University of Hawaii, 2680 Woodlawn Drive, 
Honolulu, HI 96822, USA}

 \author{Lidia N. Makarova}
 \affil{Special Astrophysical Observatory, the Russian Academy of Sciences,Nizhnij Arkhyz,    Karachai-Cherkessian Republic,
   Russia 369167} 

\author{Dmitry I. Makarov}
\affil{Special Astrophysical Observatory, the Russian Academy of Sciences,Nizhnij Arkhyz,    Karachai-Cherkessian Republic,
   Russia 369167}

\author{Luca Rizzi}
\affil{W. M. Keck Observatory, 65-1120 Mamalahoa Hwy, Kamuela, HI 9
6743, USA}

\begin{abstract}

   Hubble Space Telescope Advanced Camera for Surveys has been used to
determine accurate distances for the spiral galaxy NGC~2683 and 12 other galaxies
in a zone of the "local velocity anomaly" from measurements of the luminosities of
the brightest red giant branch stars.  These galaxies lie in the Leo Spur, the 
nearest filament beyond our home Local Sheet.  The new, accurate distance
measurements confirm that galaxies along the Leo Spur are more distant than
expected from uniform cosmic expansion, hence have large peculiar velocities
toward us.  The motions are generally explained by a previously published model 
that posits that the Local Sheet is descending at 259 km/s toward the south supergalactic
pole due to expansion of the Local Void and being attracted at 185 km/s toward
the Virgo Cluster.  With the standard $\Lambda$CDM cosmology an empty void expands 
at 16 km/s/Mpc so a motion of 259 km/s requires the Local Void to be impressively large 
and empty.  Small residuals from the published model can be attributed to an upward push toward 
the north supergalactic pole by expansion of the Gemini-Leo Void below the 
Leo Spur.  The Leo Spur is sparsely populated but among its constituents there 
are two associations that contain only dwarf galaxies.

\end{abstract}


\maketitle

\section{Introduction}

  The anisotropic nature of gravitational collapse leads to formation of the
cosmic large-scale structure, whose basic elements are walls ("pancakes"),
filaments, and clusters ("nodes") on their intersection (Zeldovich 1970,
Peebles 1980, Shandarin et al. 2012). According to the results of N-body
simulations, about 77\% of the volume fraction of the Universe is occupied
by cosmic depressions (voids), while about 50\% of the total mass is concentrated
within filaments (Cautun et al. 2014).

  Computer simulations predict that filaments and
walls can move relative to each other with velocities of several hundred
km/s. Bulk motions of the same order are expected along
the filaments towards rich clusters (Shandarin \& Klypin 1984, Shandarin \&
Zeldovich 1989, van Haarlem \& van de Weygaert 1993). Observational confirmation
of these predictions are extremely scarce because they require precise
distances to many galaxies measured independently from their radial velocities.

  To date, we know that the nearest and the most completely studied flat
configuration of a dozen nearby groups, called the "Local Sheet", is a rather
cold structure with a characteristic dispersion of peculiar (non-Hubble)
velocities $\sigma_v\sim40$ km/s (Karachentsev et al. 2003, Tully et al. 2008).
Recently, Brinckmann et al. (2014) found similar sheets with
$\sigma_v\sim100$ km/s near the rich Coma Cluster.

  According to Tully et al. (2008, 2013), the Local Sheet as a whole is
moving toward a nearby cloud of galaxies called the Leo Spur in the 
"Nearby Galaxies Atlas" (Tully \& Fisher 1987) with a velocity of
$\sim320$ km/s. The basic components of this bulk motion of the Local
Sheet are its infall towards the nearby Virgo Cluster, as well as motion
away from the Local Void, with velocities $\sim190$ km/s and $\sim260$
km/s, respectively. Addition of the vectors leads to appearance on the sky
of the zone of a "Local Velocity Anomaly"  in Gemini-Cancer-Lynx
constellations, a subject of discussion since the 1980's
(Faber \& Burstein 1988, Tully 1988a).

In supergalactic coordinates, the galaxies in the Local Sheet lie in the 
equatorial band, SGZ = 0, stretching toward the Virgo Cluster.  The Local Void lies 
toward the positive pole.  The Leo Spur lies in a plane roughly parallel to
the Local Sheet toward the negative pole.  Below the Leo Spur is the
"Gemini-Leo" Void,  No. 27 in the
list of nearby volumes completely devoid of galaxies (Elyiv et al. 2013).
The void center is near RA = $7\fh8$, Dec. = $+24^{\circ}$ at a
distance of 18.4 Mpc and its radius is 7.5 Mpc, or 24$^{\circ}$.
The Local Sheet and the Leo Spur are moving toward each other,
squeezed between two inflating voids.
Accurate measurements of the relative motions provide insight into
the importance of the juxtaposed voids.

   The Leo Spur contains several dozen galaxies with radial velocities less
than 500 km/s, but accurate distances have only recently become available
for a subset. 
The program "The Geometry and Kinematics of
the Local Volume" with Hubble Space Telescope, GO 12546 \& 13442 (PI:
R.B.Tully) supplemented by program GO 12658 (PI: J. Cannon), 
has gone a long way toward augmenting the galaxy census in this area by
facilitating distance determinations with the tip of the red giant branch (TRGB)
method. Below, we present a summary of observational data on distances and
peculiar velocities of galaxies in the sky region RA =[$5^h15^m - 11^h30^m$],
Dec =[$-8^{\circ} - +56^{\circ}$], with velocities in the Local Group rest
frame below $V_{LG}=500$ km/s. The compilation includes new accurate
distance measurements for the luminous spiral galaxy NGC~2683 and for 12
dwarf galaxies.
       
\section {ACS HST observations and data processing}
      
  We have observed 10 galaxies with Advanced Camera
for Surveys (ACS) during HST Cycle 21 (proposal 13442) 
that are relevant to the present discussion.  
Between October 25, 2013 and April 21, 2014 we obtained
1000s F606W and 1000s F814W images of each galaxy using
ACS/WFC with exposures split to eliminate cosmic ray contamination.
In addition we have reduced and analyzed images obtained for
three galaxies in the relevant region observed during cycle 19
with proposal 12658 (PI: J. Cannon).  Additional material from HST program 12546
and earlier was discussed by Jacobs et al. (2009) and is included in the {\it Cosmicflows-2} data release (Tully et al. 2013)
and made available at the Extragalactic Distance Database.\footnote{http://edd.ifa.hawaii.edu catalog CMDs/TRGB}

The new images were obtained from the STScI archive,
as processed according to the standard ACS pipeline.
The stellar photometry was performed using the ACS module of DOLPHOT
(http://americano.dolphinsim.com/dolphot), the successor to HSTPHOT
(Dolphin 2000), using the recommended recipe and parameters.  In brief,
the process involves the following steps.  First, pixels that are flagged as bad
or saturated in the data quality images were marked in the data images.
Second, pixel area maps were applied to restore the correct count rates.
Finally, the photometry was run. In order to be reported, a star had to
be recovered with S/N of at least four in both filters, 
$\vert sharp \vert \le 0.3$, be unsaturated and relatively
clean of bad pixels (such that the DOLPHOT error flag is zero) in both
filters. These restrictions reject non-stellar and
blended objects. At the high Galactic latitudes of the targets, foreground
stars from the Milky Way are insignificant contaminants.
All the observed galaxies are resolved into individual stars including
those on the
red giant branch (RGB), allowing a distance to be measured with the tip of 
the red giant branch (TRGB) method.

  The TRGB is determined by a maximum likelihood analysis monitored by 
recovery of artificial stars (TRGBtool software, Makarov et al. 2006). 
Artificial stars with a wide range of known magnitudes and colors are 
imposed over the frame in numbers relative to the density of the real stars and 
recovered (or not) with the standard analysis procedures
to determine both photometric errors and completeness in the crowded field 
environments. The maximum likelihood procedure evaluates the luminosity 
function of stars with colors consistent with the red giant branch after 
compensating for completeness and assesses power law fits to the 
distributions above and below a break identified with the TRGB.
The slope of the power law faintward of the TRGB break is expected to be 
approximately 0.3 on a magnitude scale after correction for completeness.  
If the RGB is sufficiently observed to well below the tip then the slope can 
be a free parameter within a restricted range but in the current cases, with 
distances approaching the effective observational limits, the slope of the 
luminosity function fit below the TRGB is set to the expected value of 0.3.
Galactic extinction is taken from Schlafly \& Finkbeiner (2011).

  The greatest potential for serious error with a TRGB measurement
arises with confusion between the intermediate age asymptotic giant branch 
(AGB) and the RGB. Stars on the AGB, that are burning both helium and hydrogen 
in shells, closely parallel and overlap the RGB on a CMD but rise as much as a 
magnitude brighter.  Their peak brightness, dependent on age and metallicity, 
can be misinterpreted as the TRGB. AGB stars have intermediate ages of 1-10 Gyr 
although they are only in sufficient quantity to be confusing at the lower 
end of that age range (Jacobs et al. 2011). A general strategy that we employ 
is clipping of the area of the HST image to avoid regions of young and 
intermediate age stars and associated obscuration (as well as regions beyond the target dominated by background 
and foreground contaminants) in order to maximize the contrast
of the old population contributing to the RGB.
Since all the galaxies under study are located relatively far from us,
their TRGB position is close to the photometric limit. In such cases, 
the RGB luminosity function may not have the clear discontinuity necessary 
for an application of the maximum likelihood method. In such cases, the TRGBtool program uses a simple age 
detection algorithm, which can result in larger TRGB uncertainties.
Five galaxies from our sample (UGC 3600, UGC 3698, NGC 2337, UGC 3860, UGC 5288)
have RGB luminosity functions that are too smooth for the maximum likelihood calculation.
The calibration of the absolute value of the TRGB including
a small color term has been described by Rizzi et al. (2007).
The RGB is redder for older or more metal rich populations but
galaxies inevitably have old and metal poor components, resulting in
reasonable stability of tip magnitudes in the F814W band.   The F814W band
is favorable because the TRGB luminosity in this band is minimally dependent on age and metallicity effects.
Images, color-magnitude diagrams, photometry tables, TRGB measurements, and 
distance determinations are made available for the newly observed galaxies 
at http://edd.ifa.hawaii.edu 
by selecting the catalog CMDs/TRGB (Jacobs et al. 2009).

\section{TRGB distances to thirteen galaxies}

\begin{figure*}
\includegraphics[width=5.4cm]{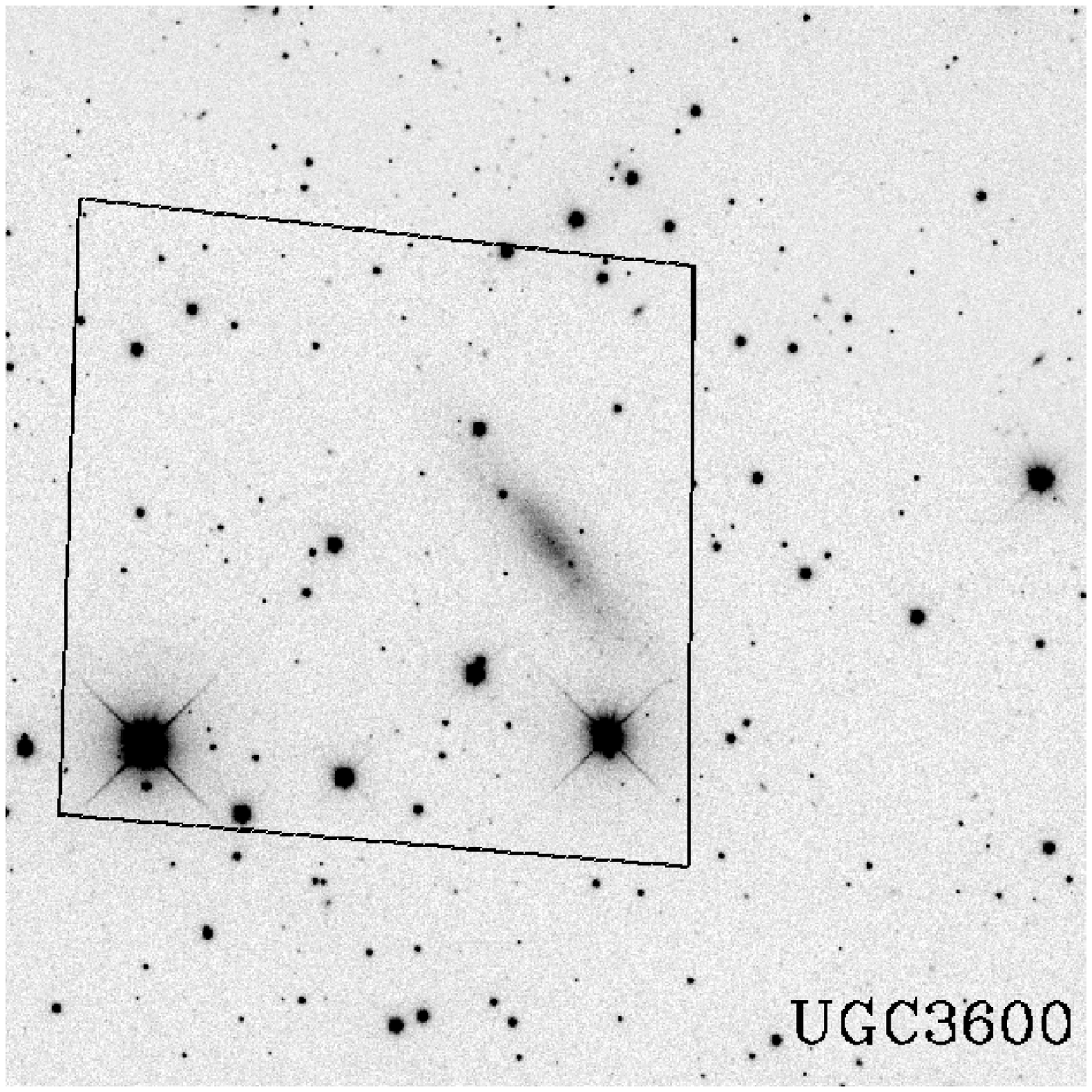}
\includegraphics[width=5.4cm]{u3698imdss1.ps}
\includegraphics[width=5.4cm]{n2337imadss.ps}
\includegraphics[width=5.4cm]{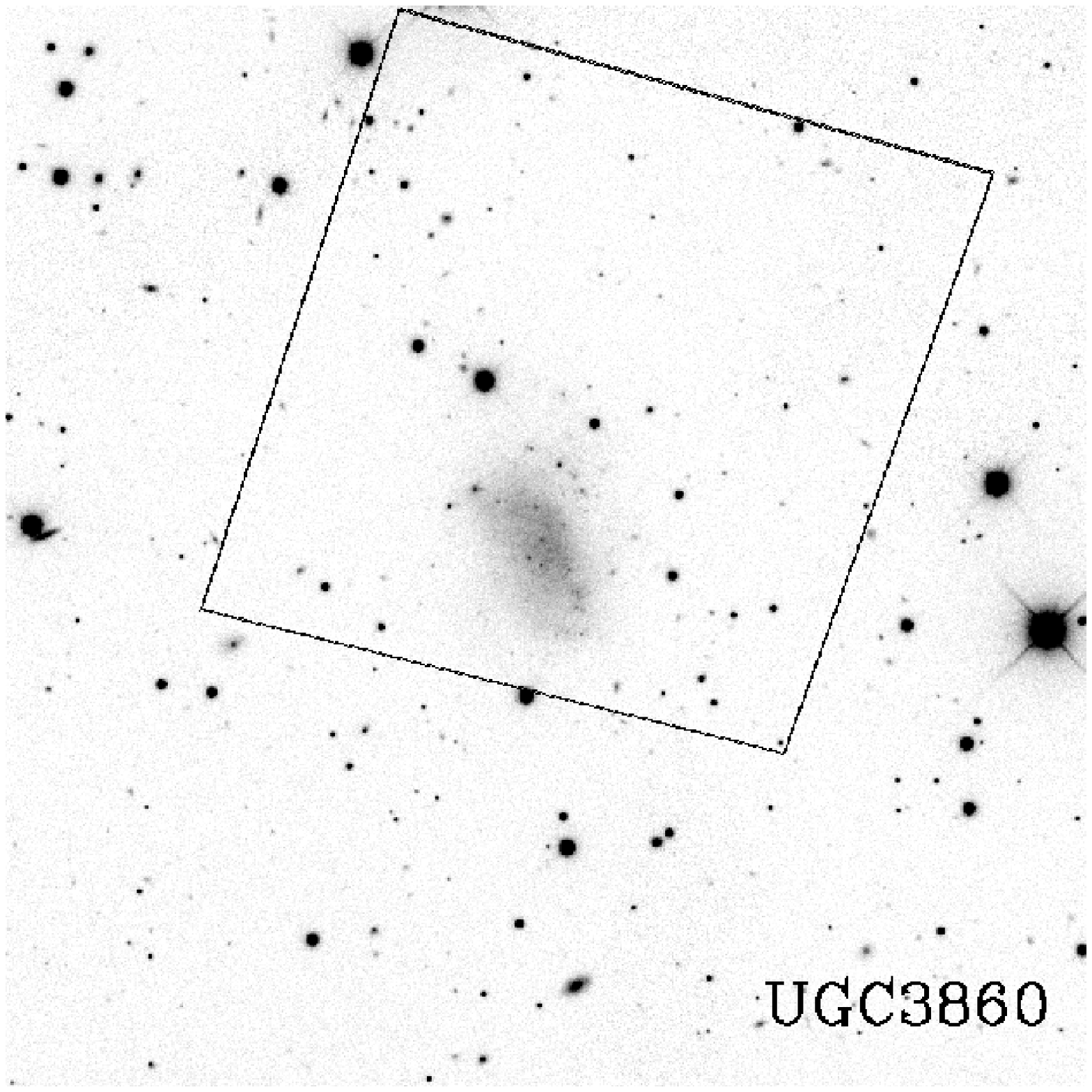}
\includegraphics[width=5.4cm]{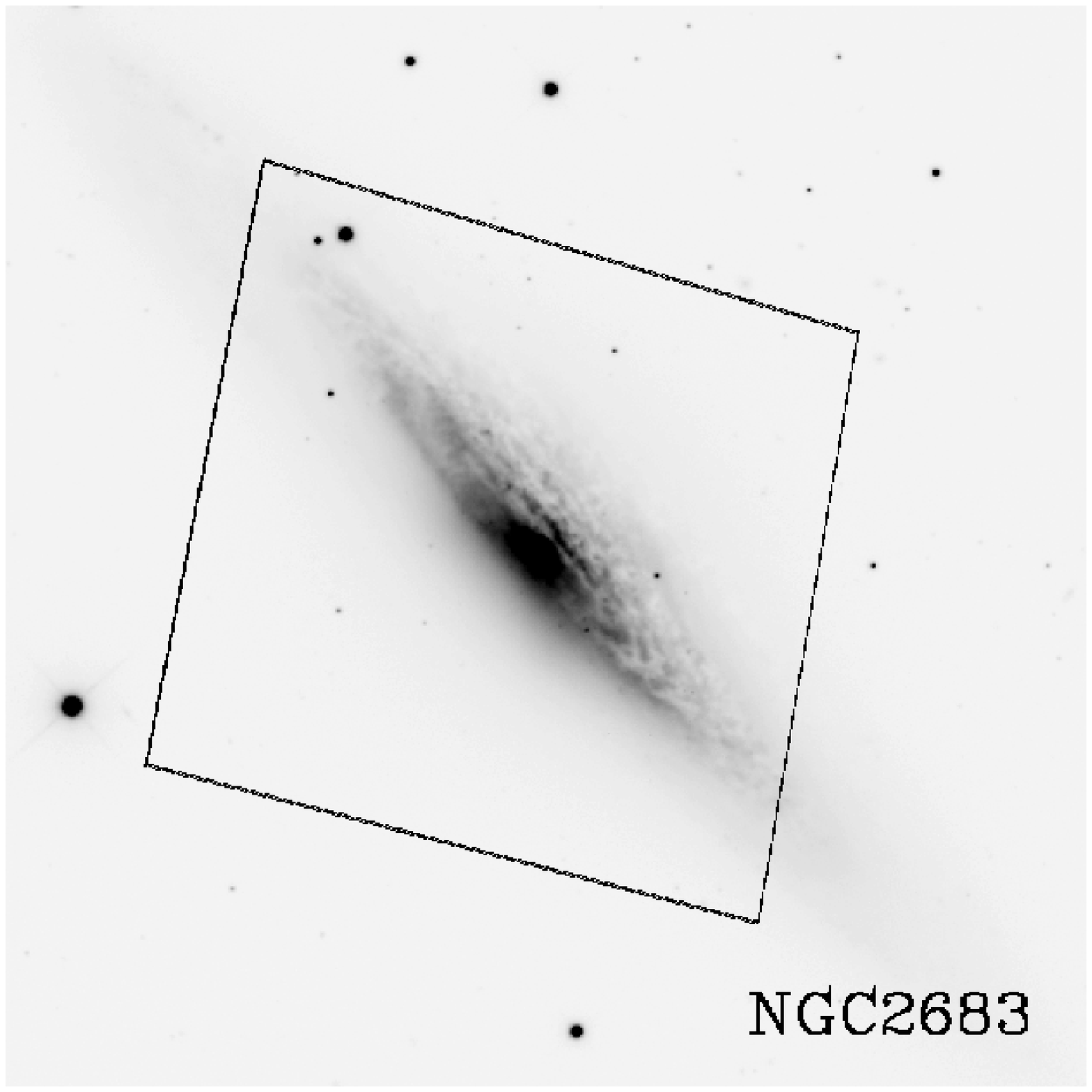}
\includegraphics[width=5.4cm]{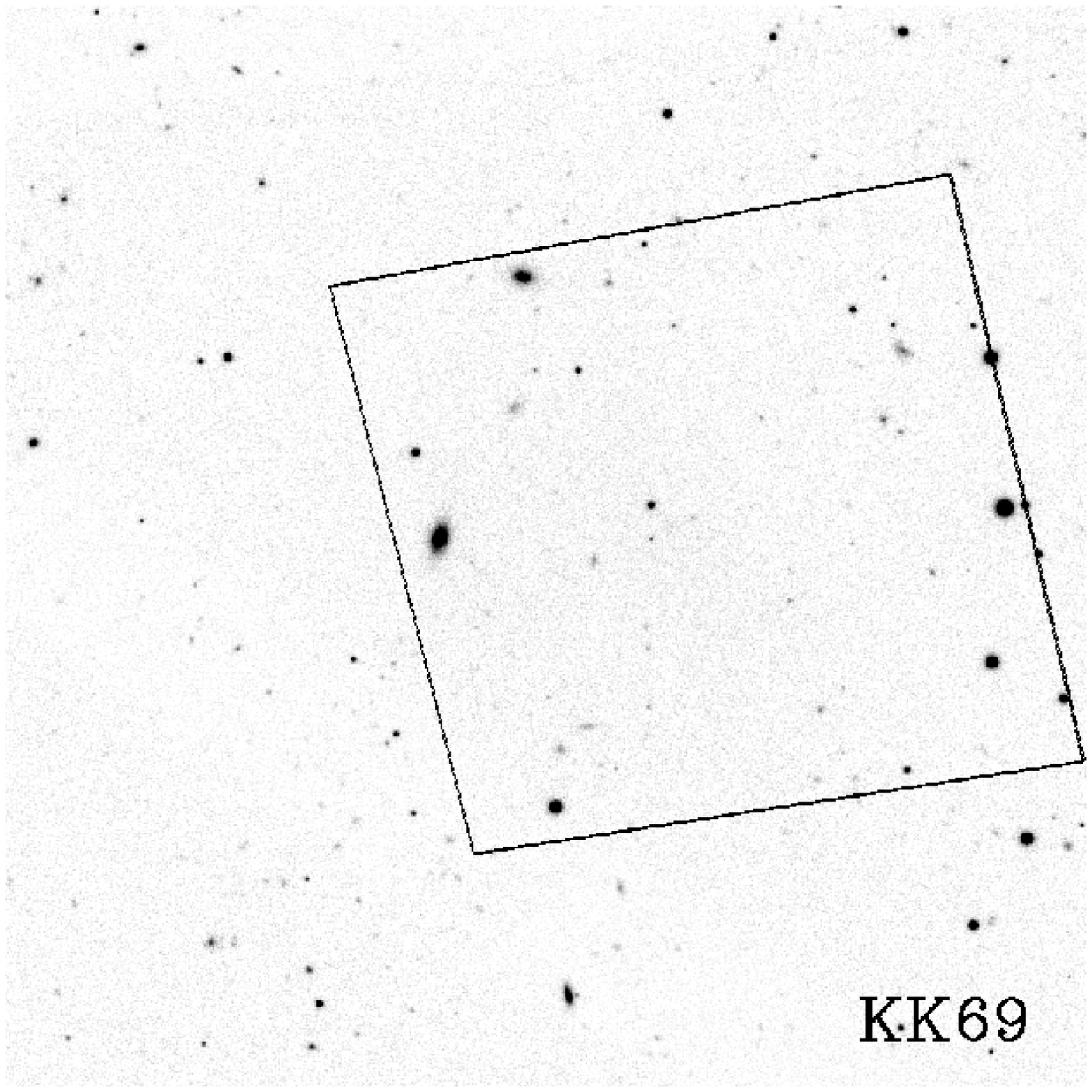}
\includegraphics[width=5.4cm]{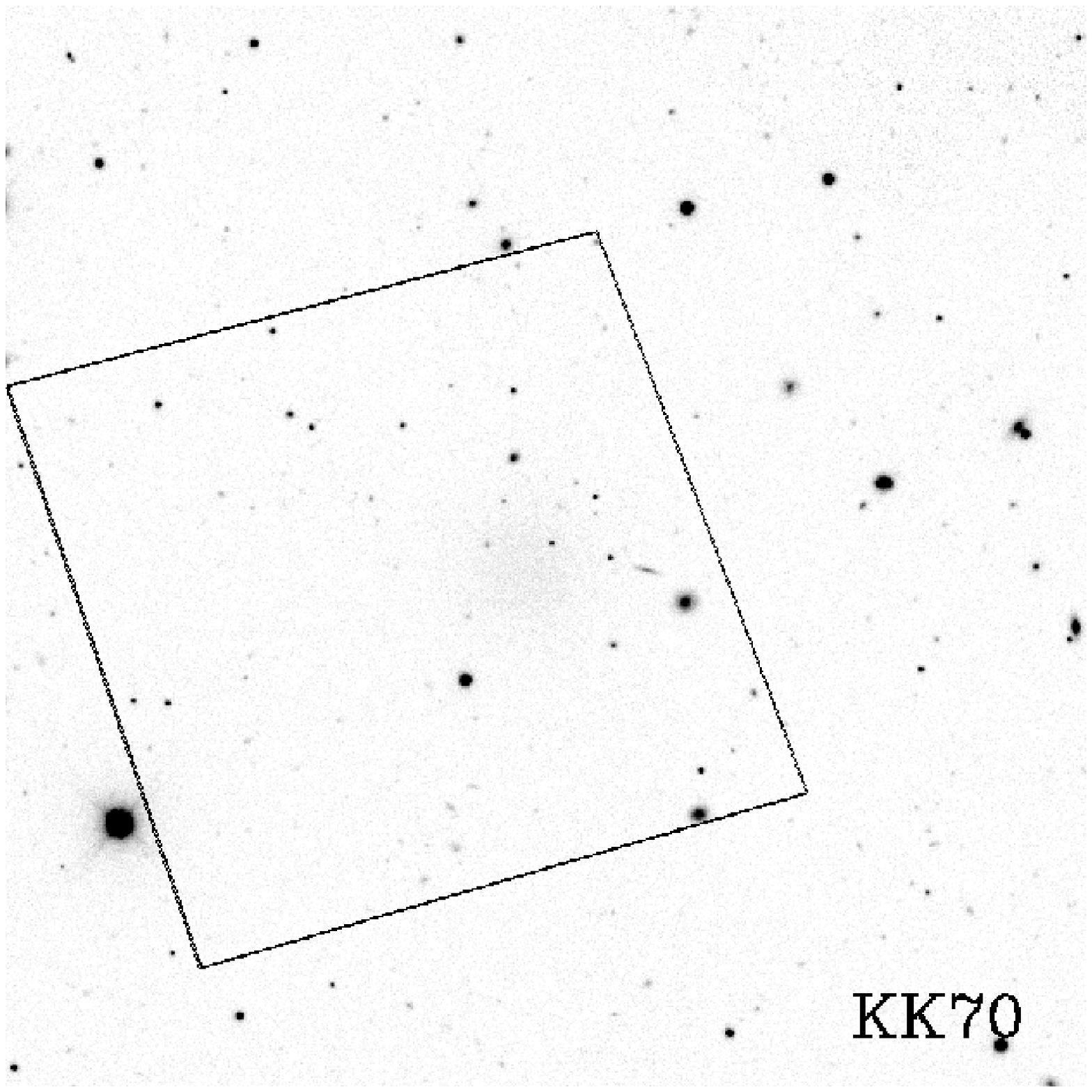}
\includegraphics[width=5.4cm]{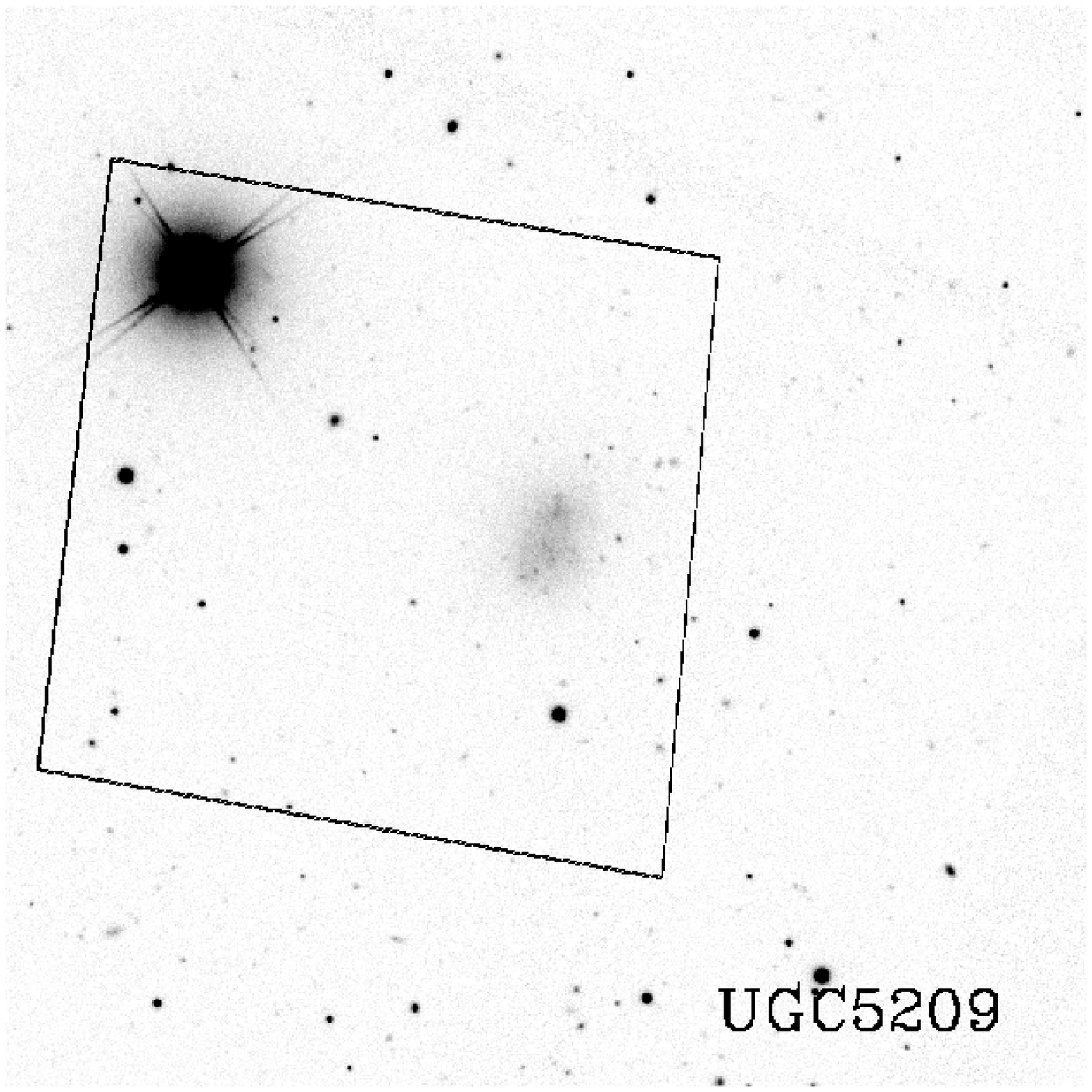}
\includegraphics[width=5.4cm]{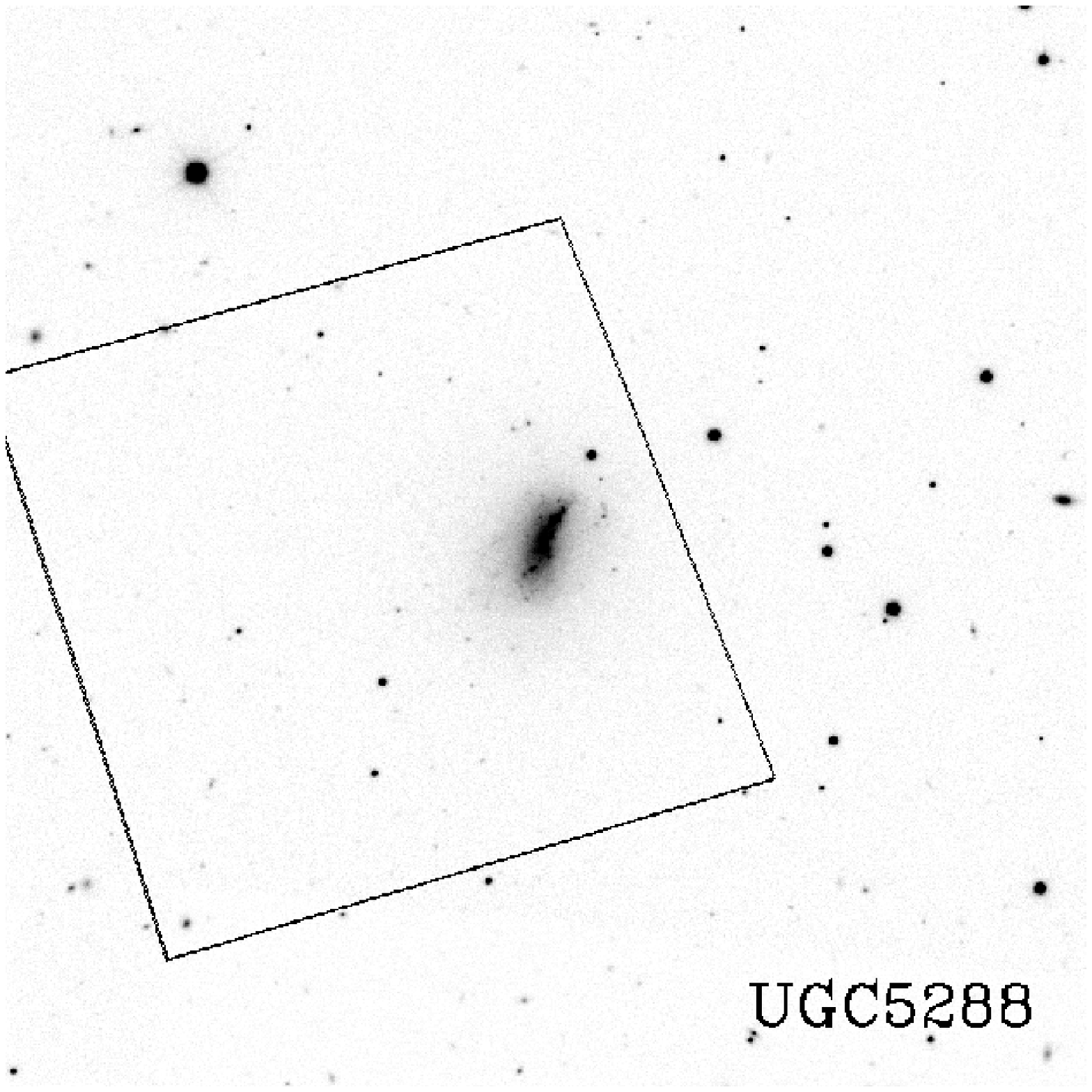}
\includegraphics[width=5.4cm]{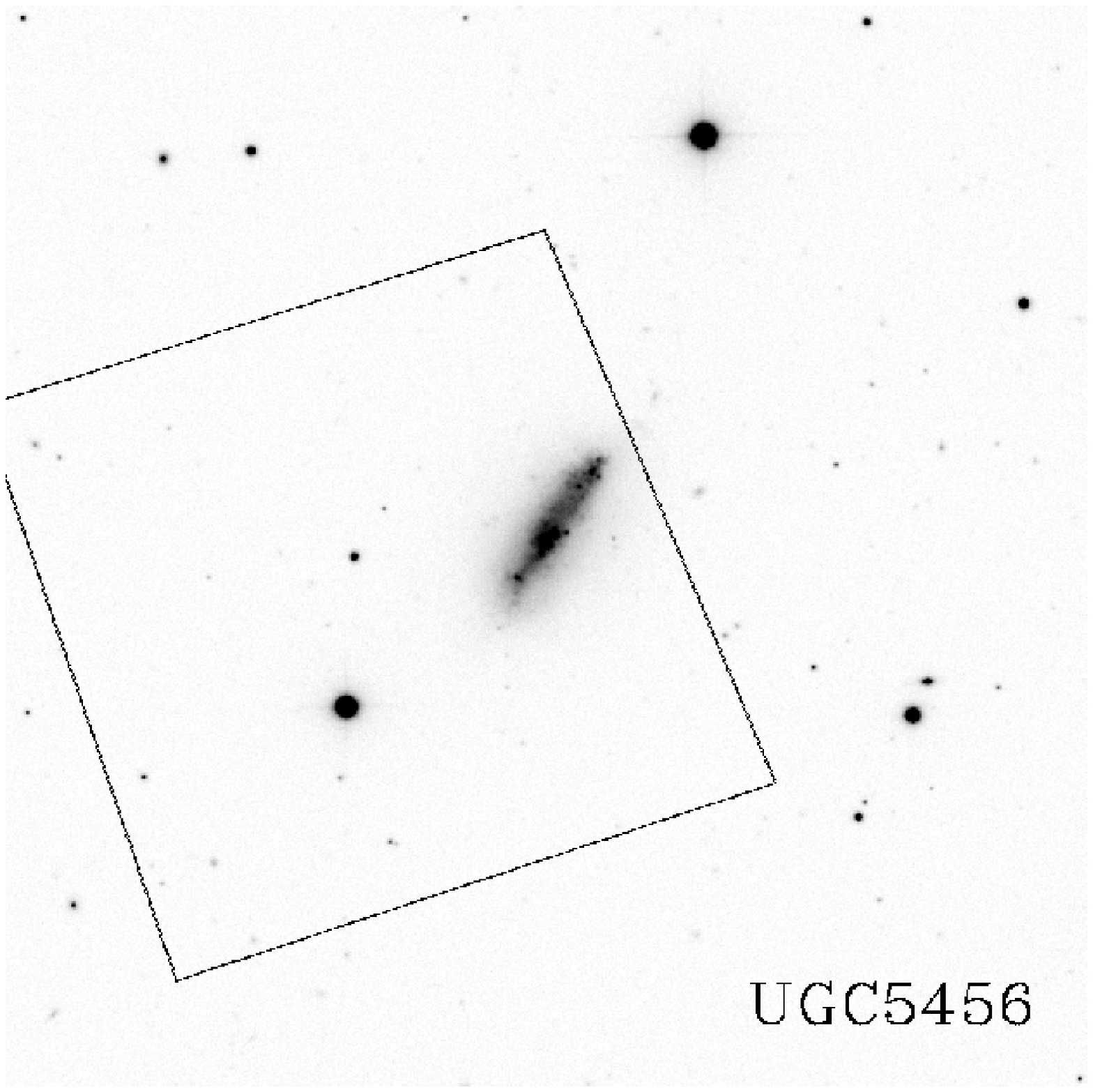}
 \caption{Sloan Digital Sky Survey images of target galaxies. Each field
       has a size of 6 by 6 arcminutes. North is up and East is left.
       The HST ACS footprints are superimposed.}
       \label{footprints}
       \end{figure*}

\begin{figure*}
\includegraphics[width=5.4cm]{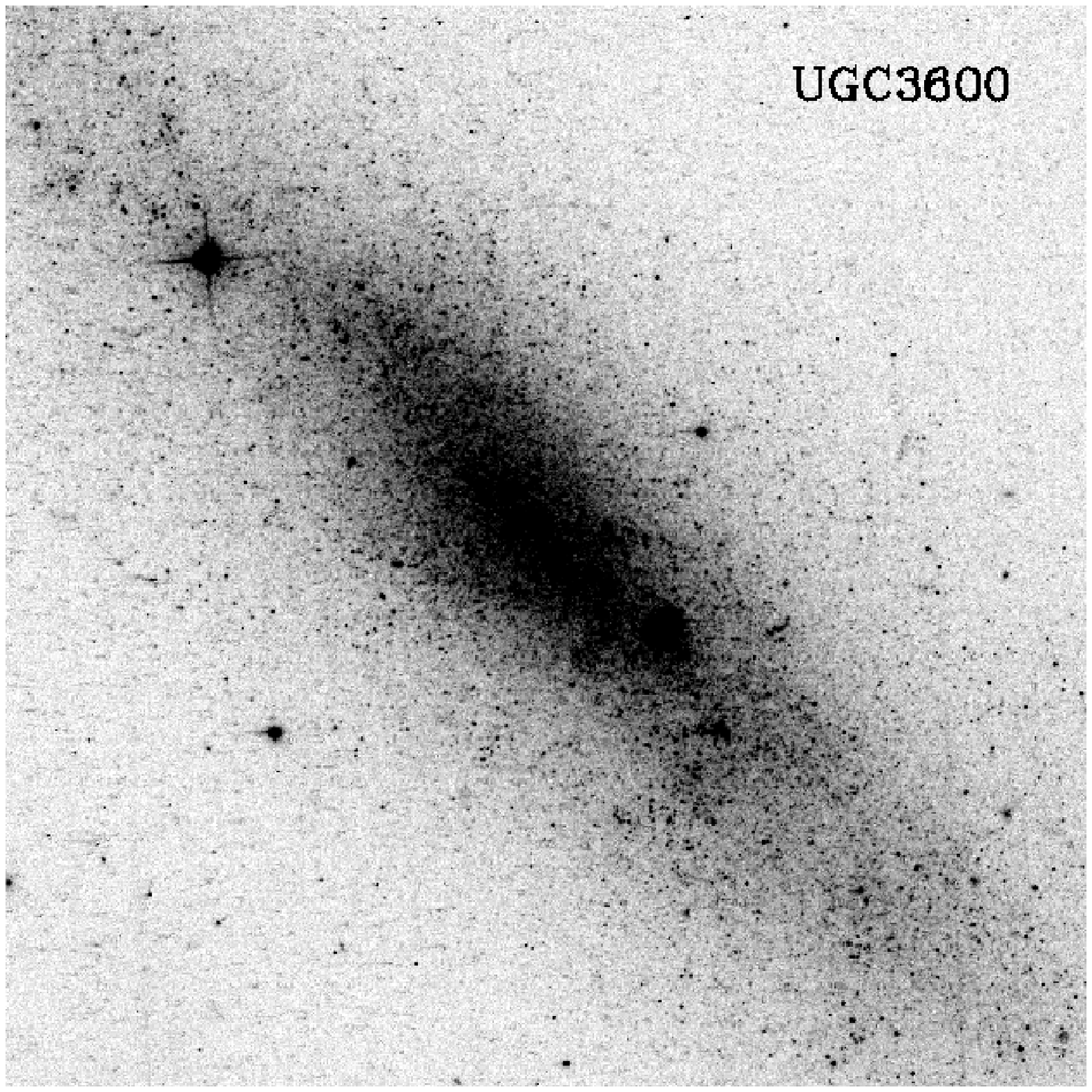}
\includegraphics[width=5.4cm]{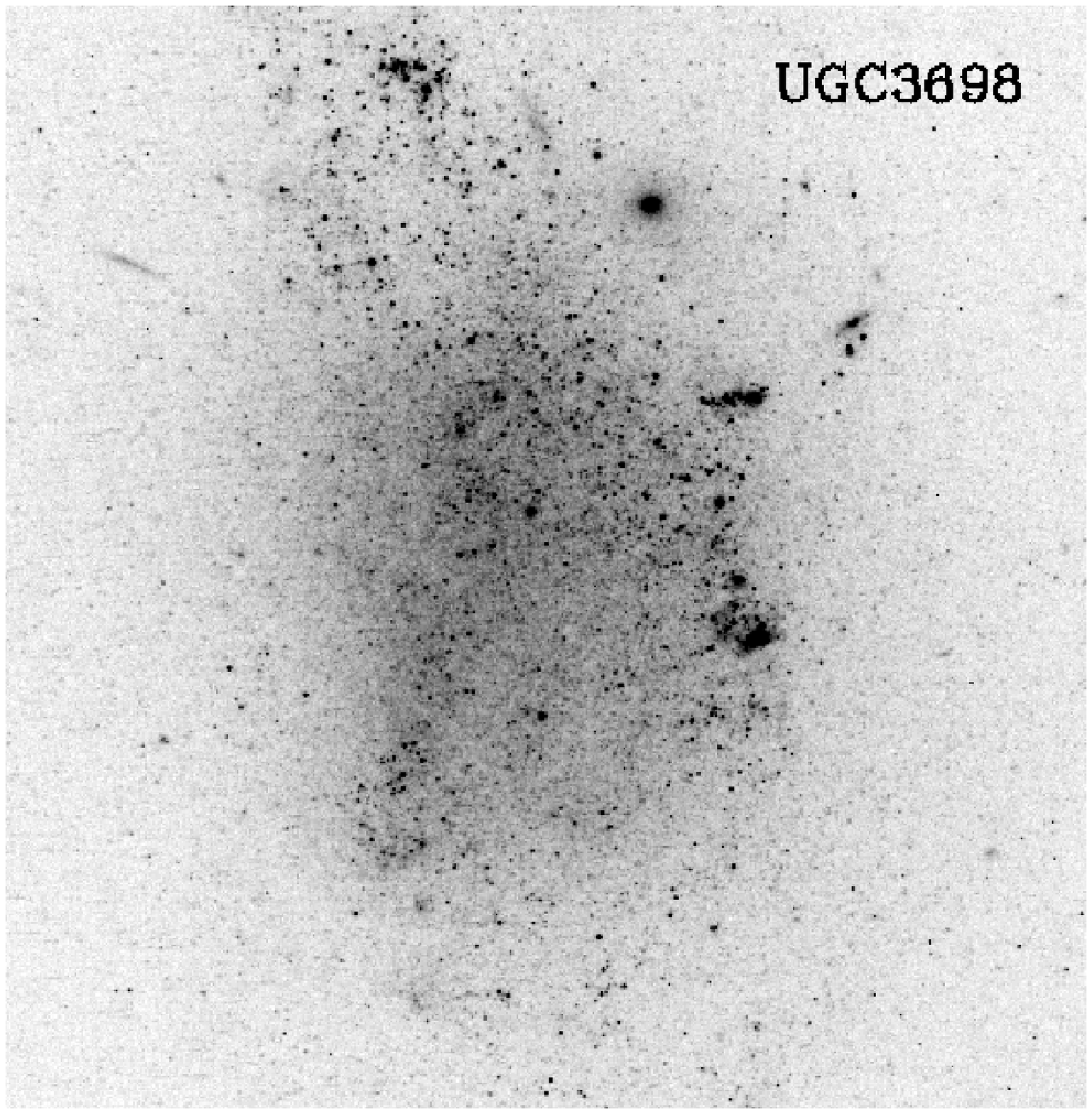}
\includegraphics[width=5.4cm]{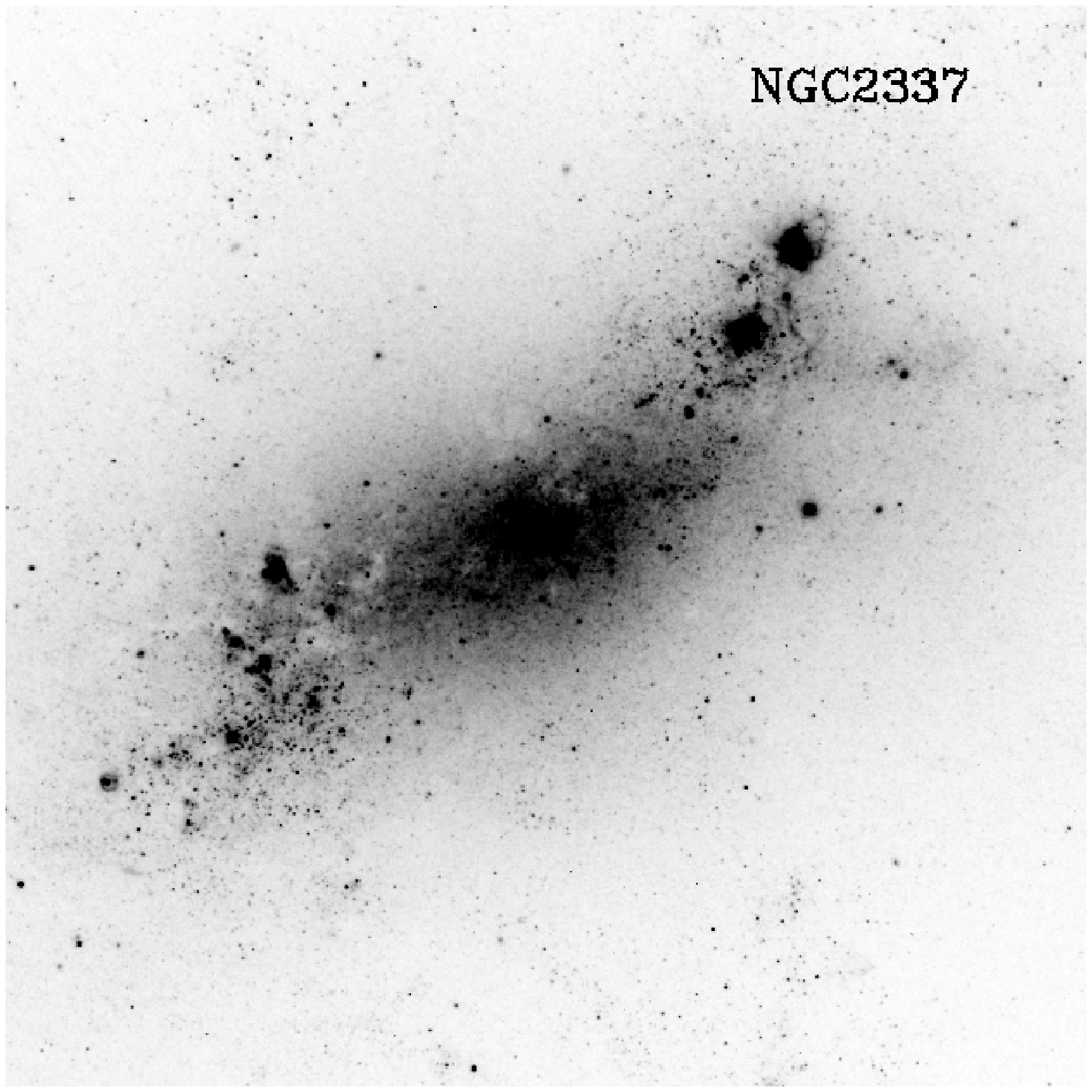}
\includegraphics[width=5.4cm]{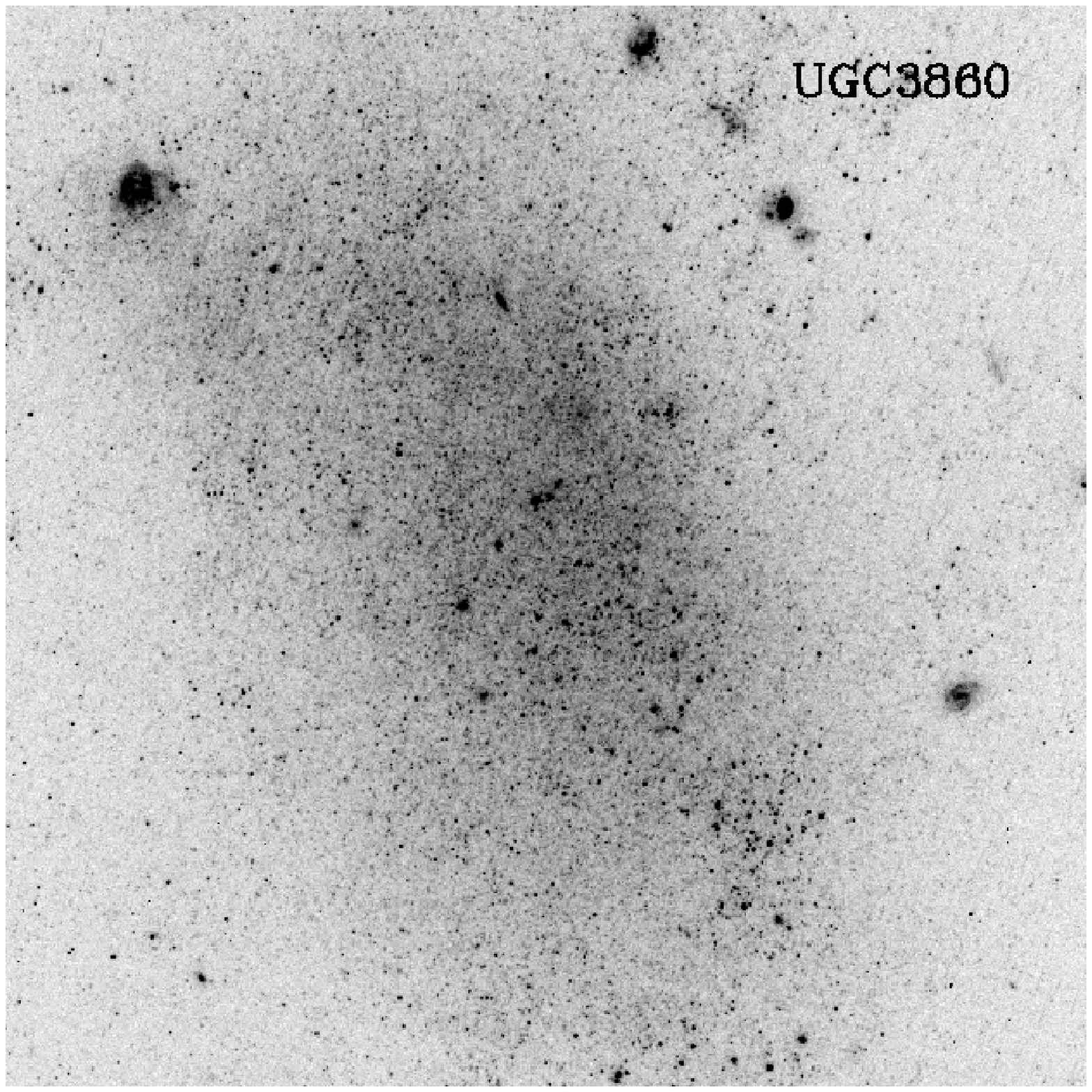}
\includegraphics[width=5.4cm]{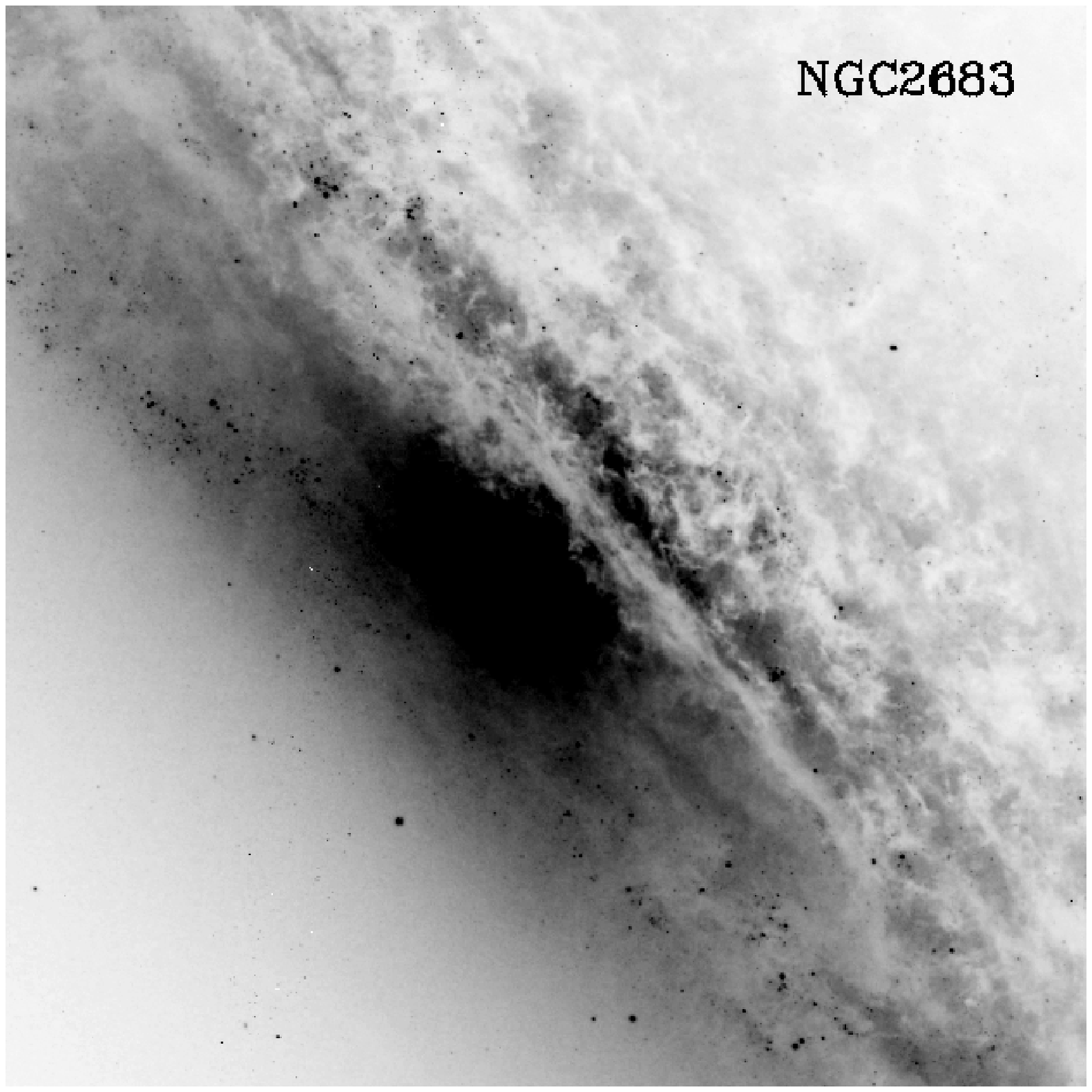}
\includegraphics[width=5.4cm]{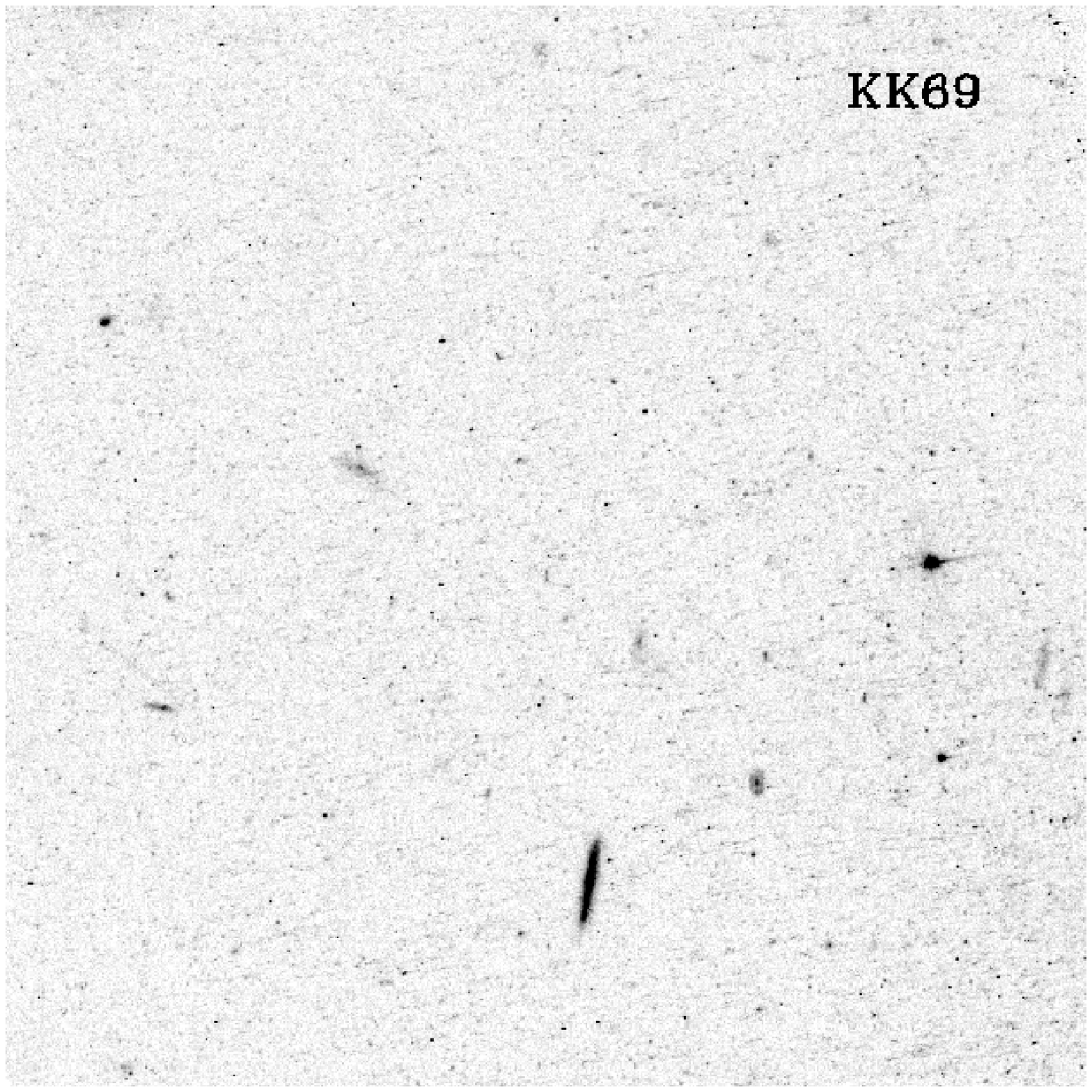}
\includegraphics[width=5.4cm]{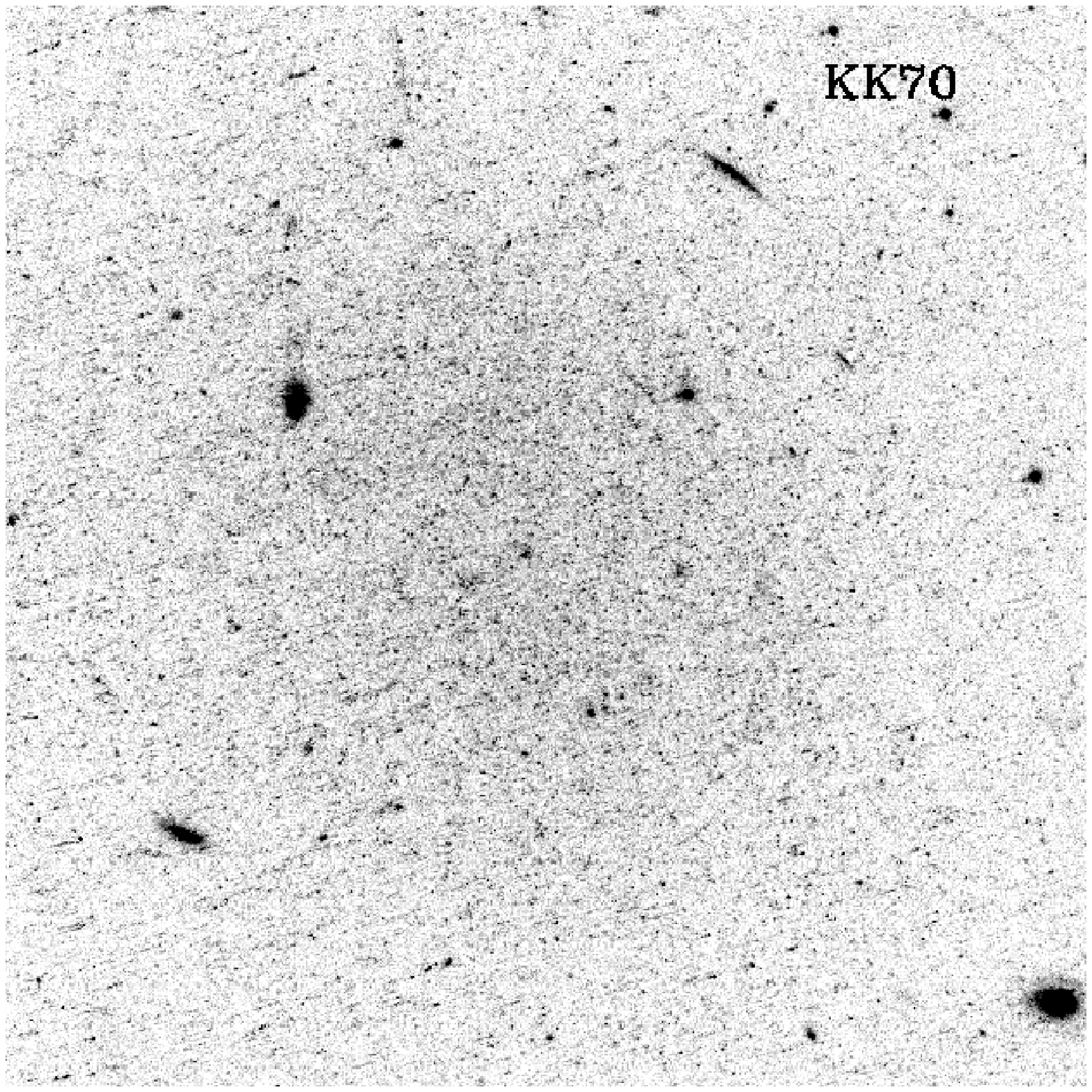}
\includegraphics[width=5.4cm]{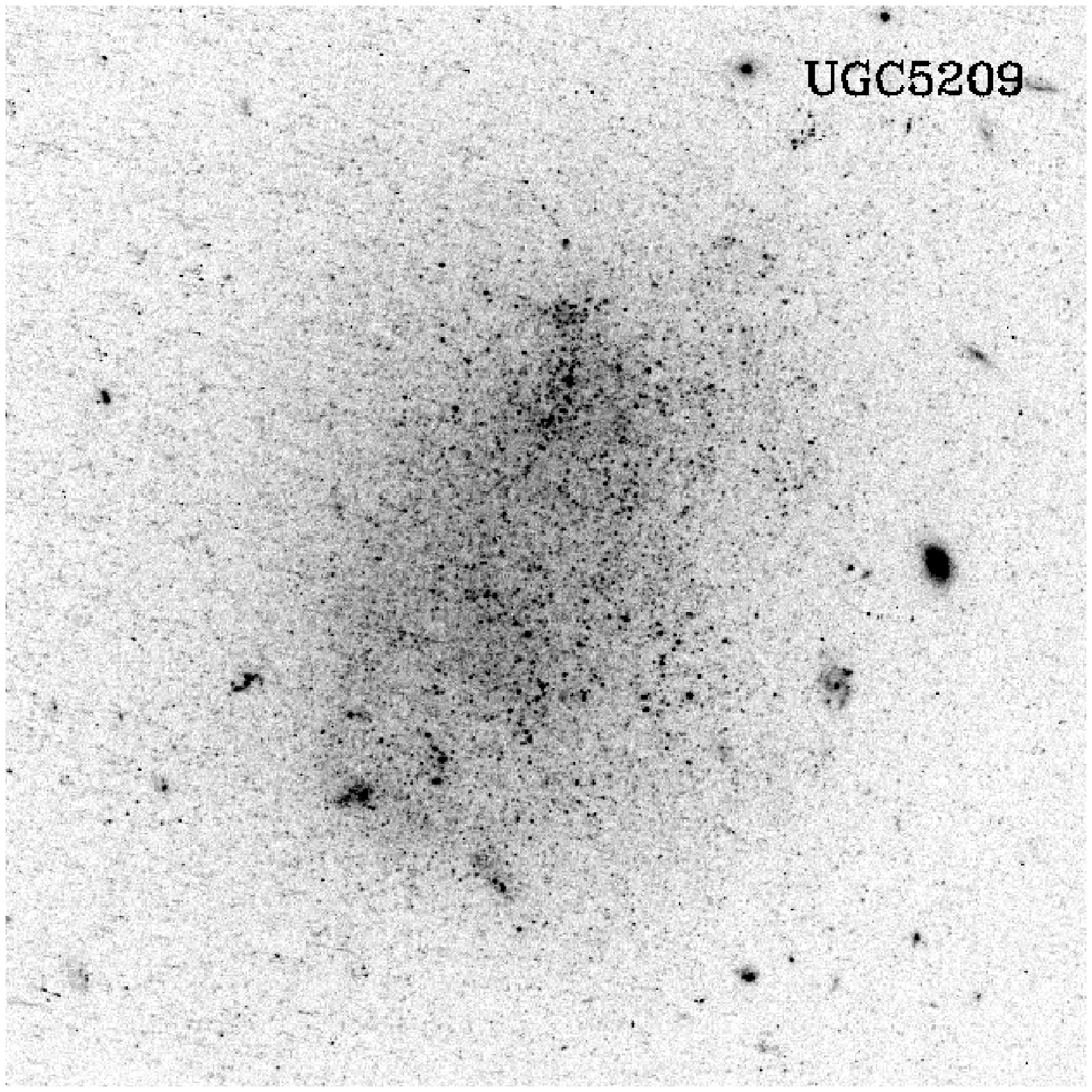}
\includegraphics[width=5.4cm]{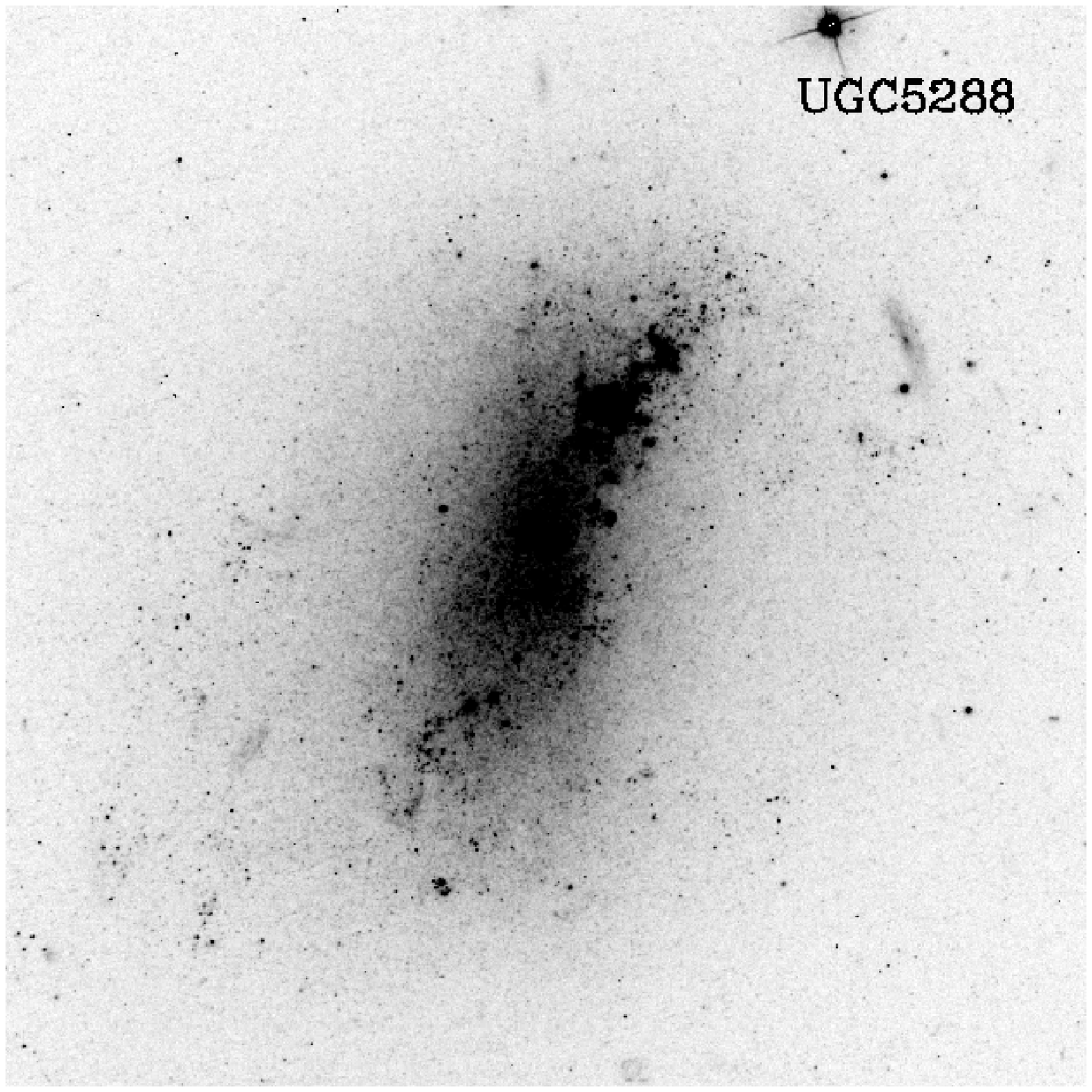}
\includegraphics[width=5.4cm]{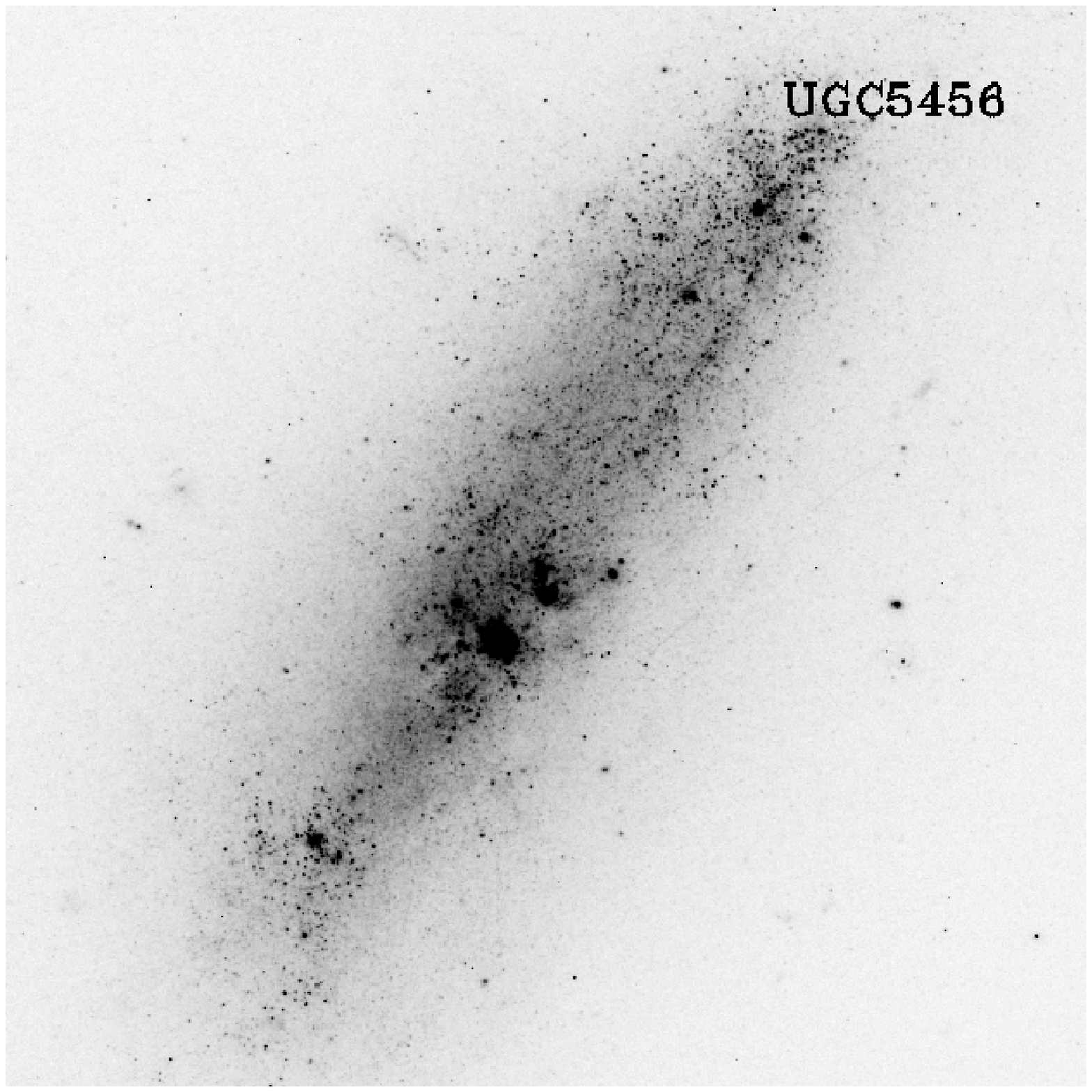}
 \caption{Mosaic of enlarged ACS F606W images of the galaxies.
       Field sizes are 1 arcminute on a side, North is up and East is left.}
       \label{images}
       \end{figure*}

\begin{figure*}
\includegraphics[width=4.3cm]{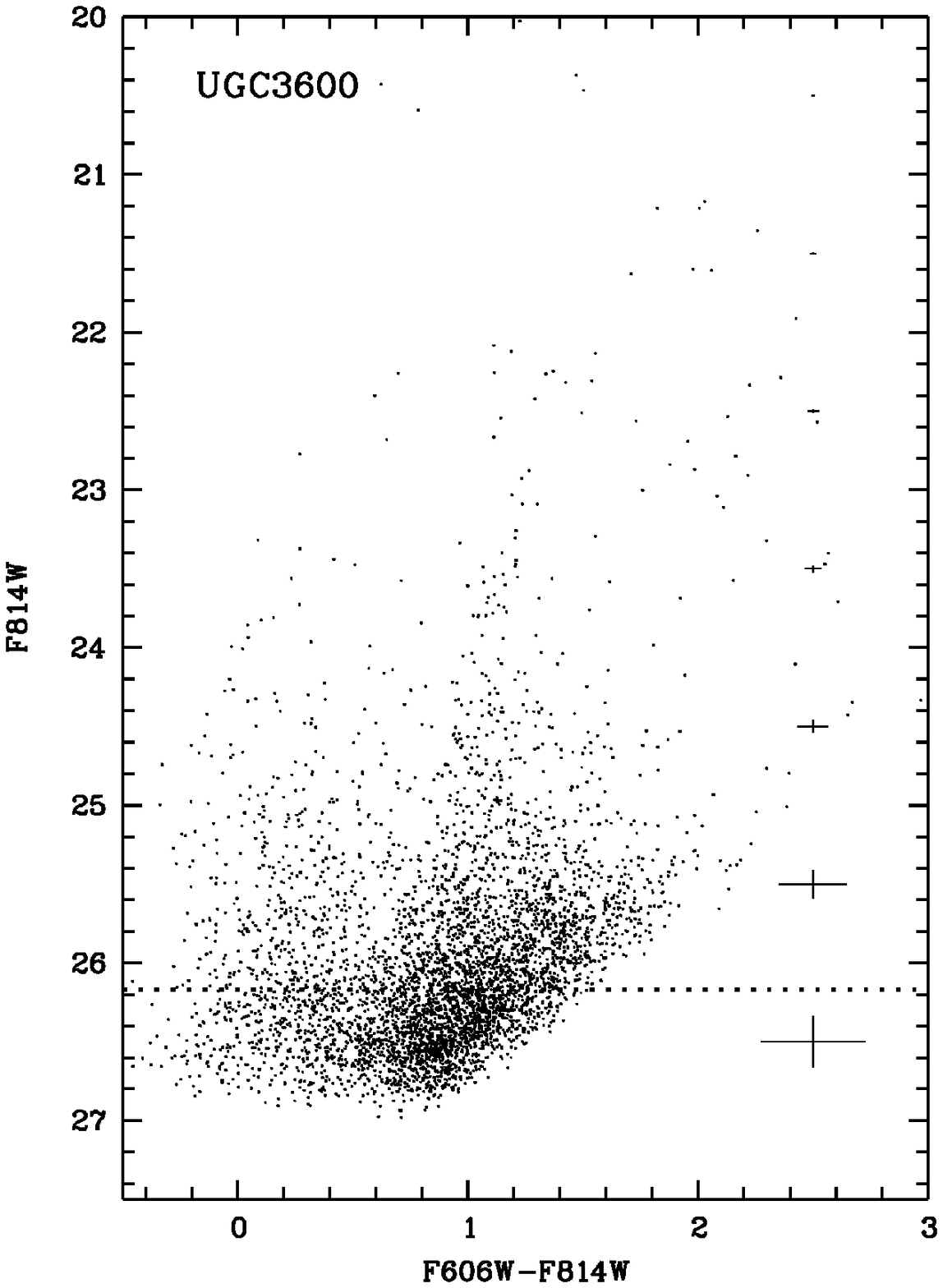}
\includegraphics[width=4.3cm]{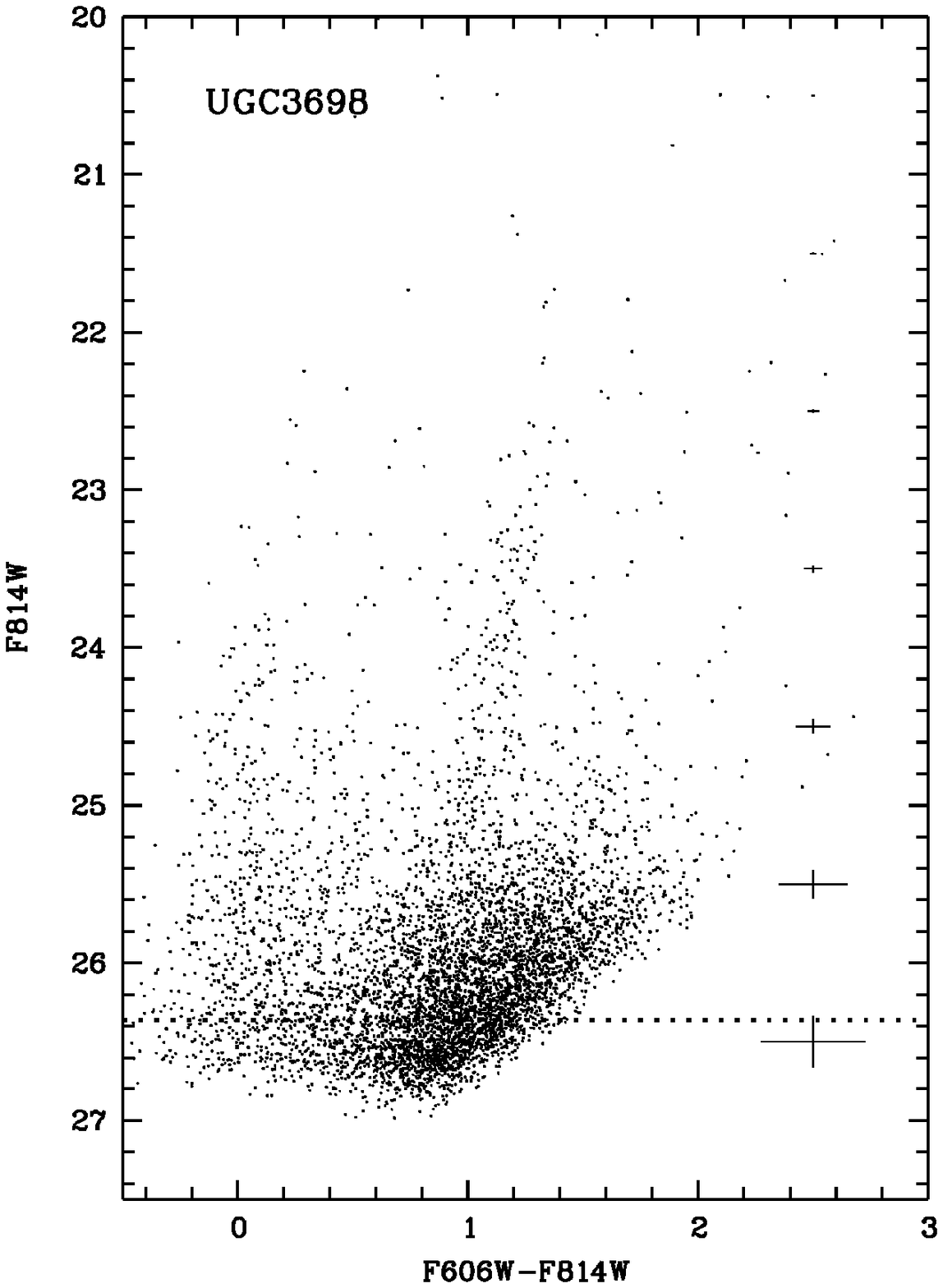}
\includegraphics[width=4.3cm]{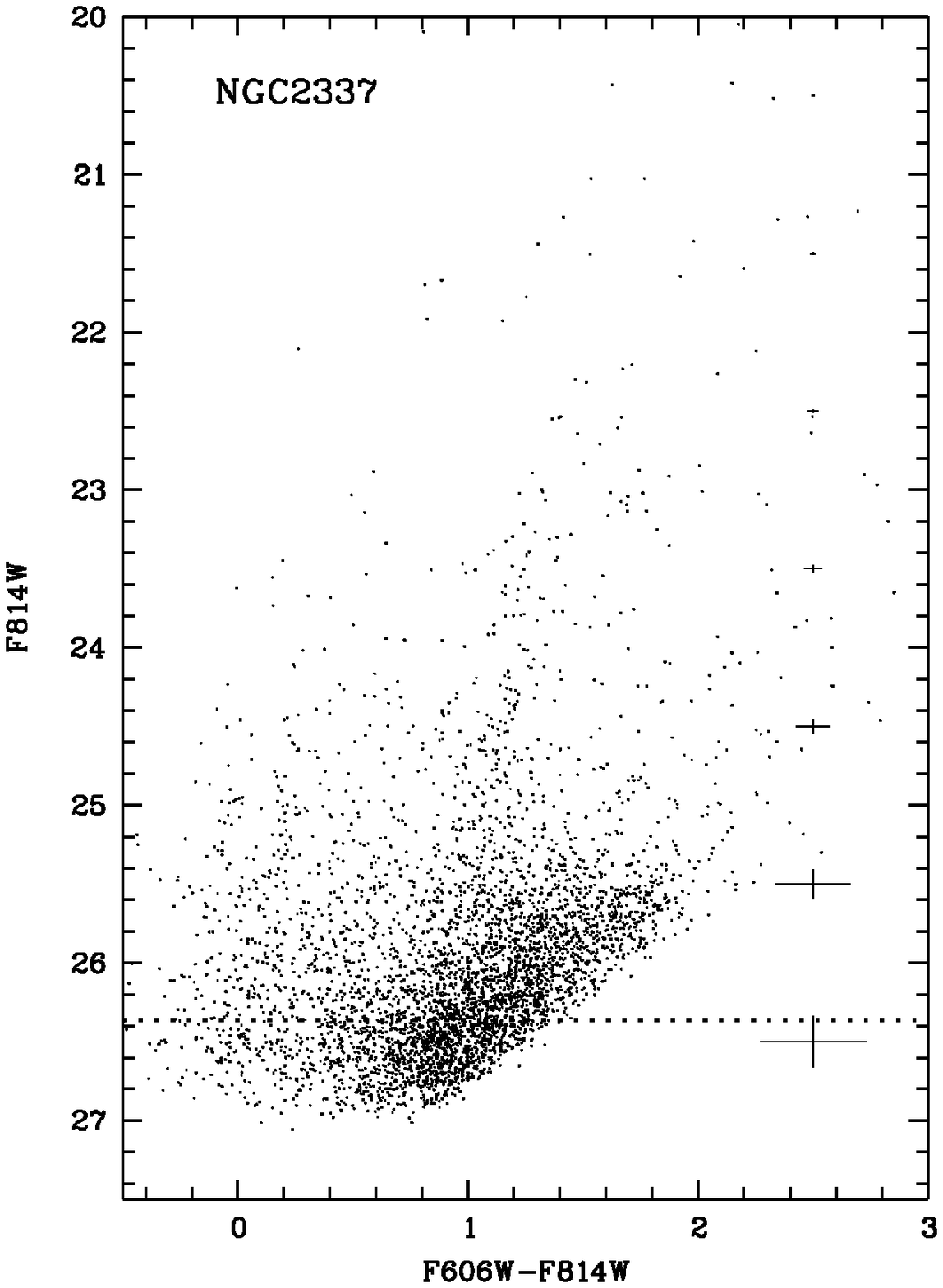}
\includegraphics[width=4.3cm]{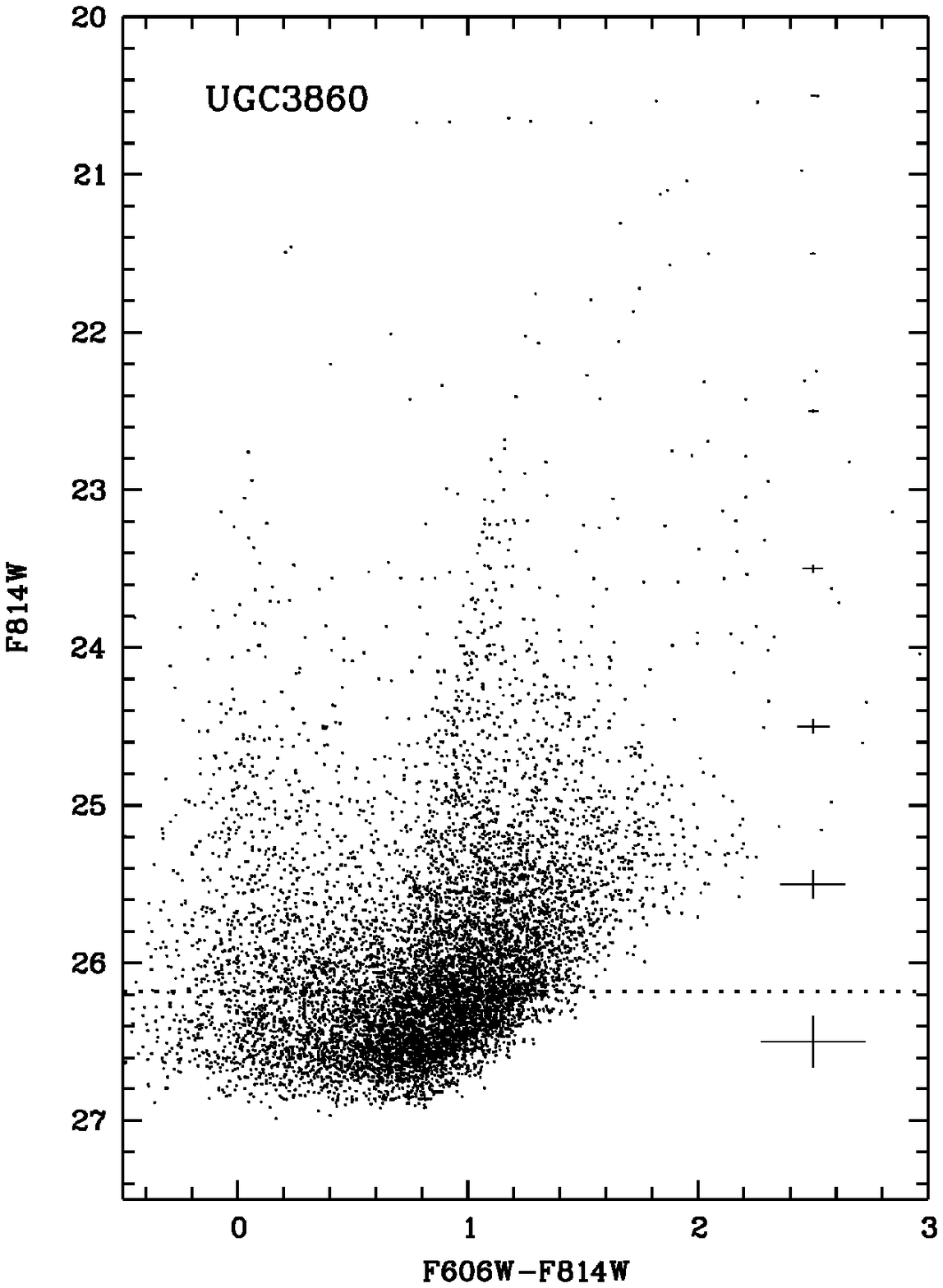}
\includegraphics[width=4.3cm]{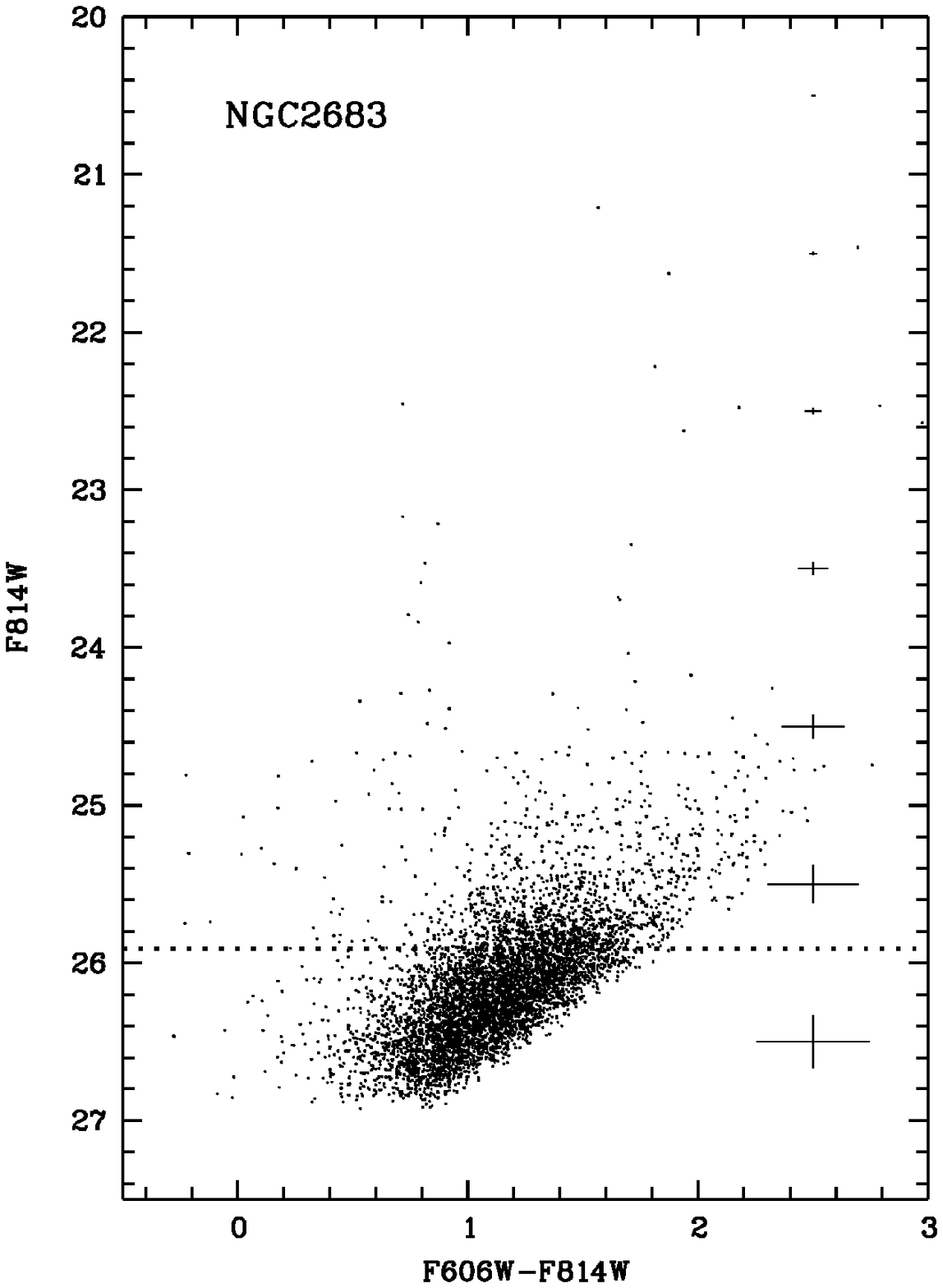}
\includegraphics[width=4.3cm]{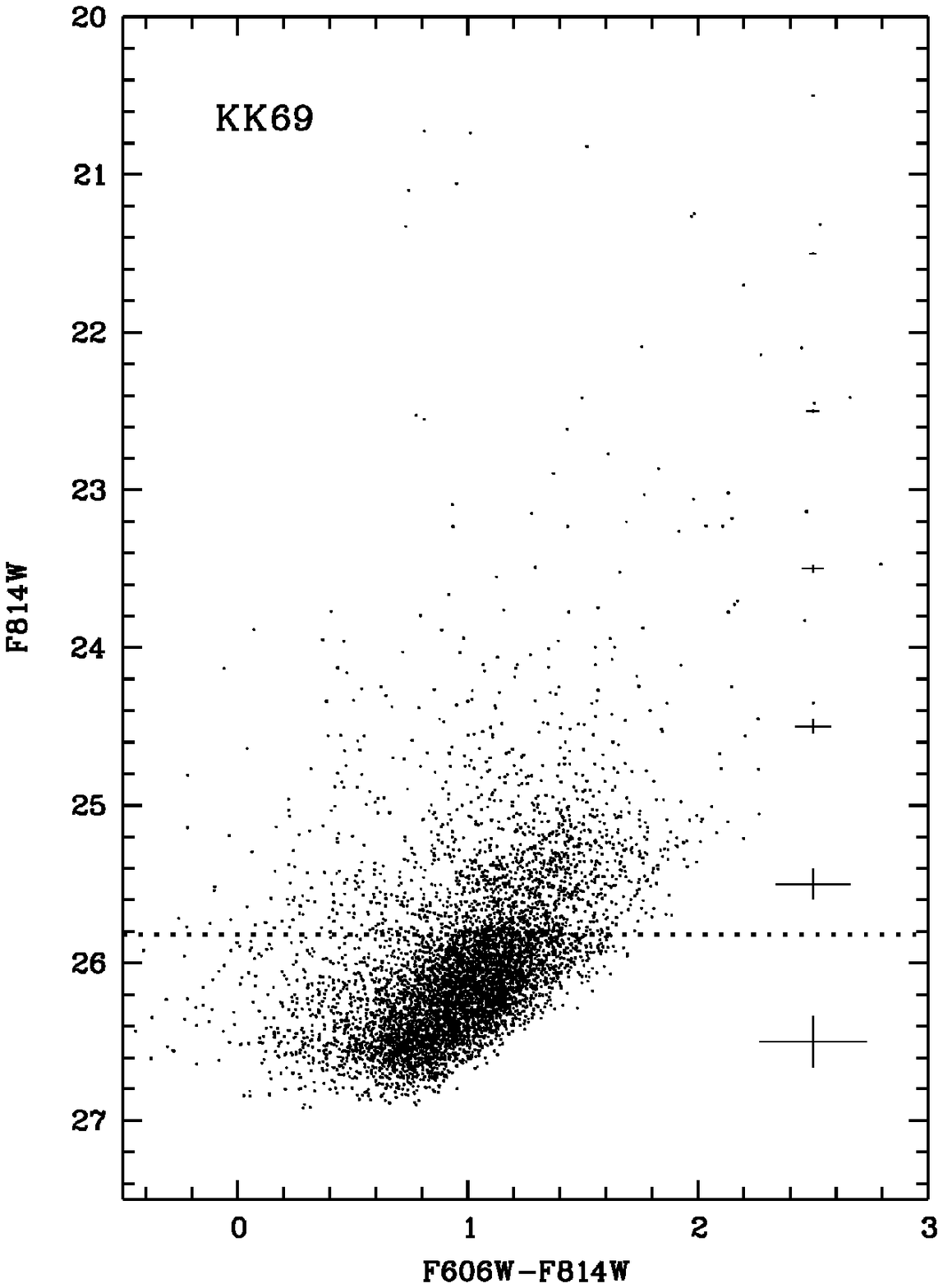}
\includegraphics[width=4.3cm]{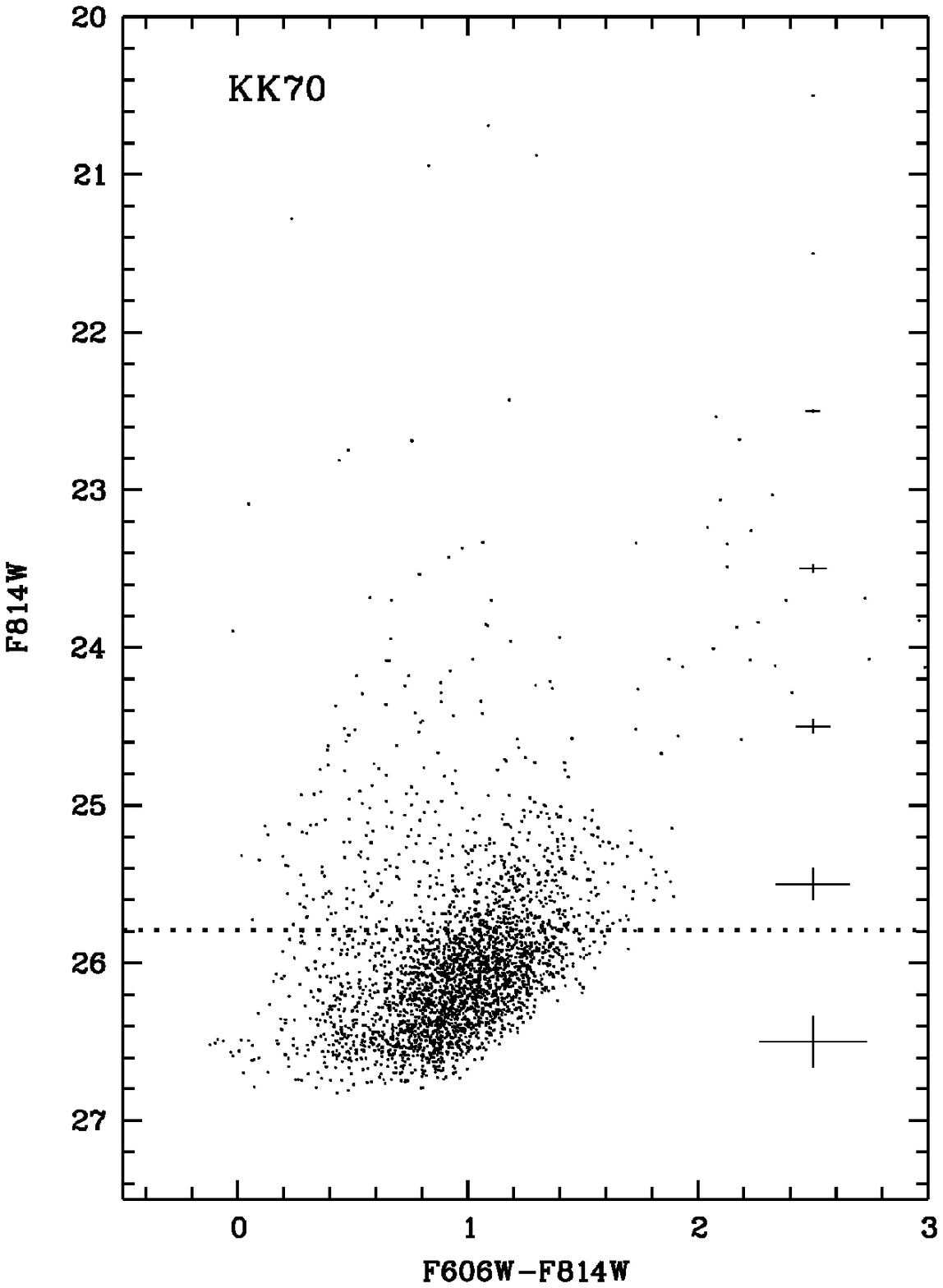}
\includegraphics[width=4.3cm]{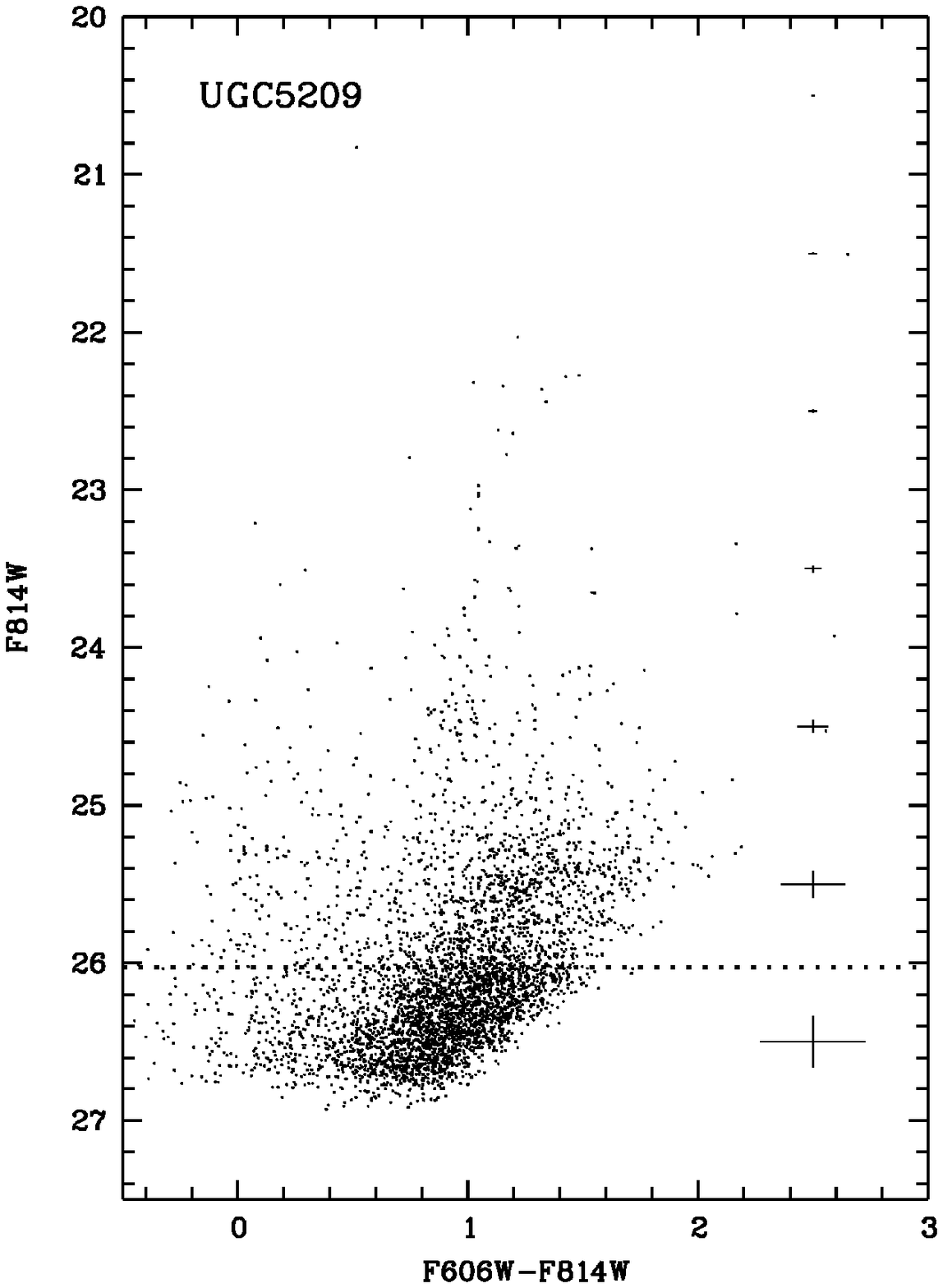}
\includegraphics[width=4.3cm]{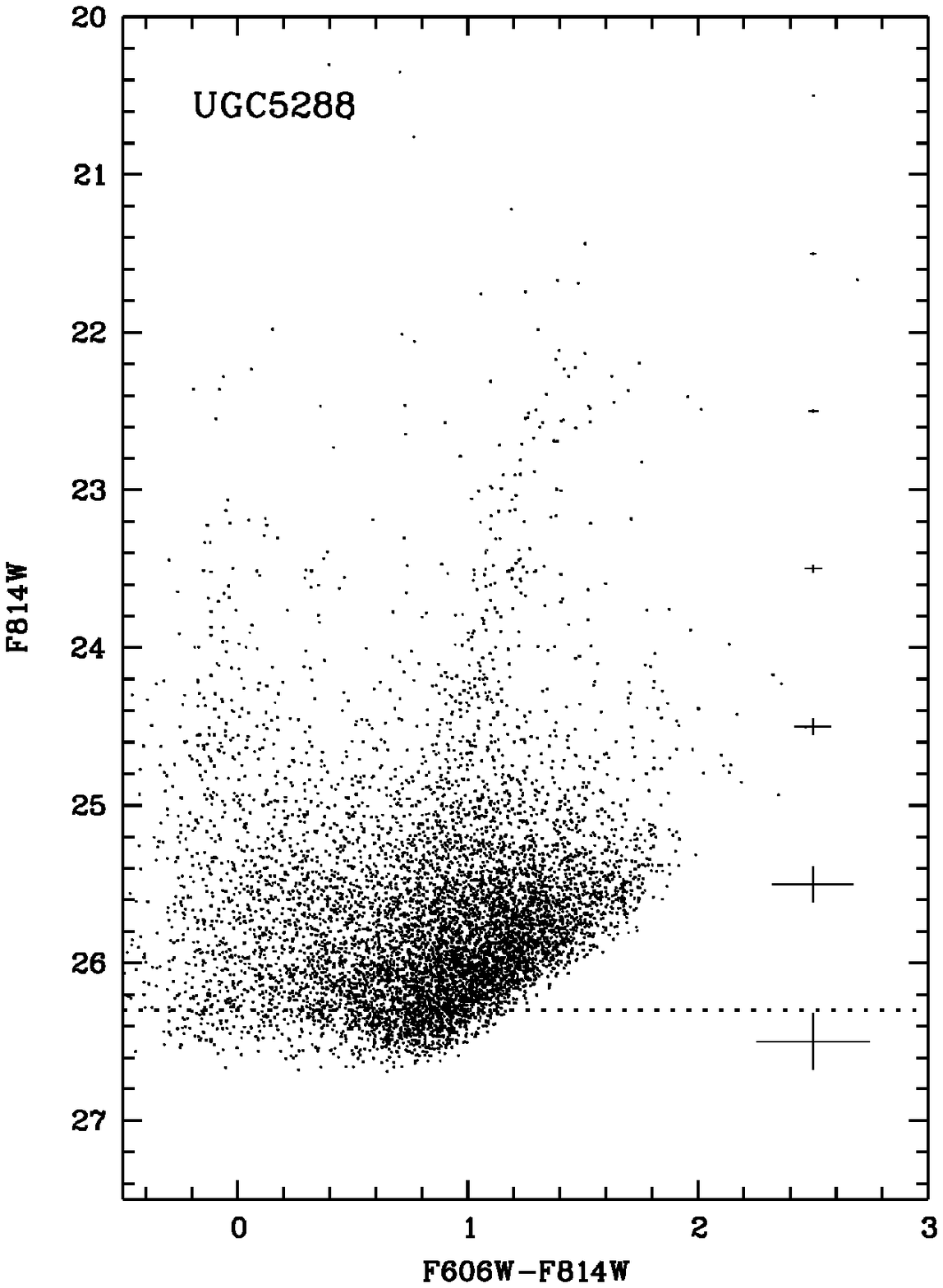}
\includegraphics[width=4.3cm]{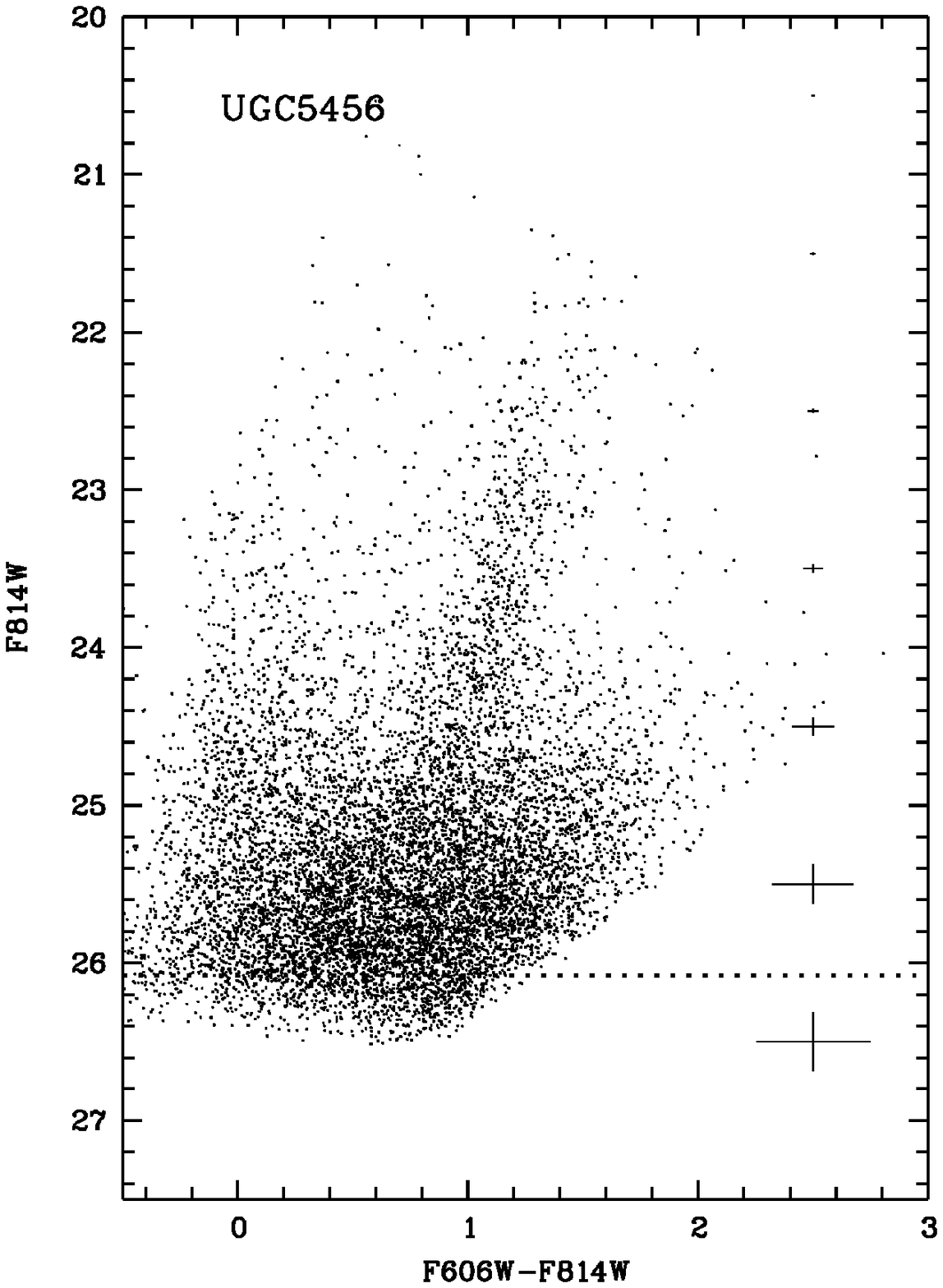}
 \caption{}
       \label{cmd}
\end{figure*}

  Images of our target galaxies taken from Sloan Digital Sky Survey
(http://www.sdss.org/) are shown in Figure \ref{footprints}. Each field has a size
of 6 by 6 arcminutes. North is up and East is left.
The ACS HST footprints are superimposed on the SDSS frames.
Figure \ref{images} is a mosaic of enlarged ACS (F606W) images of
the galaxies. The field size is 1 arcminute; North is up
and East is left. Color-magnitude diagrams (CMDs) of F814W versus
(F606W - F814W) are presented in Figure \ref{cmd}.
Table 1 provides a summary of basic parameters for the observed galaxies, 
taken largely from the Updated Nearby Galaxy Catalog (UNGC; Karachentsev et al. 
2013), as well as new TRGB and distance measurements.
There is the following information in the table columns: (1) common galaxy name, (2) Principal
Galaxies Catalog number, (3) equatorial coordinates, 
(4-5) supergalactic coordinates, (6-8) heliocentric, Local Sheet, and Local Supercluster
velocities,
(6) major diameter measured at the Holmberg 26.5 mag/square arcsec
isophote in arcmin, (7) apparent integrated B magnitude, (8) morphological 
type on the de Vaucouleurs numeric scale, (9) TRGB magnitude, (9) Galactic extinction in the
I- band (Schlafly \& Finkbeiner, 2011), (10) the linear distance, in Mpc, 
and 68\% probability error.

Two objects of low surface brightness,
KK~69 and KK~70, are close physical companions to the spiral NGC~2683 with
projected separations of 62 and 89 kpc, respectively. The other targets besides NGC~2683
are dwarf galaxies of morphological types Irr,
Im, BCD, and Sm. According to the UNGC, the mean stellar density in a sphere of
radius 1 Mpc around each of these galaxies is about 1/30th of the mean global density of
stellar matter. All the dwarfs in the vicinity of NGC~2683
manifest signs of sluggish star-formation as seen from their H-alpha
emission, as well as from their flux in the far-ultraviolet (Karachentsev \& Kaisina 2013). The average value
of the specific star-formation rate for the dwarfs is characterized by
the quantity
       
$$\langle \log(sSFR)\rangle=\langle \log(SFR/M^*)\rangle= -10.10\pm0.07 \ {\rm y
r}^{-1},$$

\noindent
coincident with the value of the Hubble parameter
$\log(H_0)=-10.14 \ {\rm yr}^{-1}$. This agreement is suggestive of
a smoldering process where a galaxy, not subject to external influences,
reproduces its observed stellar mass over the cosmological time
$H_0^{-1}=13.8$ Gyr quasi-continuously with the presently observed star-formation rate.

\section{Peculiar velocities in the Leo Spur}

It was demonstrated (Tully et al. 2008; hereafter TSK08) that galaxies in a filament to the 
south of the Local Sheet in supergalactic coordinates are more distant than
would be expected from the Hubble relation; ie, the galaxies have peculiar 
velocities toward us.  The explanation offered, based on a sample of almost
1800 distance measurements around the sky within 3000 km/s, was that the
structure that we live in, the Local Sheet, has a motion toward this southern 
filament due to the expansion of the Local Void to the supergalactic north of us.

The filament to the south has been called the Leo Spur (Tully \& Fisher 1987).
At the time of the TSK08 discussion, only a few galaxies in the Leo Spur had
distance measurements with sufficiently small uncertainties to compel the conclusion that
the Leo Spur and the Local Sheet are moving toward each other in co-moving 
coordinates.  The situation now is dramatically improved.

Our current knowledge is demonstrated in three sequences of figures.
In each sequence there are three panels.  The top panel shows a projection
normal to SGY$-$SGZ, identifying the positions of all galaxies in the
volume with distance measurements as reported either in the {\it Cosmicflows-2} 
compilation (Tully et al. 2013) or in this paper.   The other two panels show
the same volume with projections normal to SGY$-$SGX, split at SGZ=$-2$,
so the middle panel displays the Local Sheet on the supergalactic equatorial 
plane and the lower panel displays the Leo Spur below the equatorial plane.  
Colors of symbols identify the techniques used to obtain the 
distances.  The virial regions of the Virgo and Leo clusters are located with
filled circles and the domain of the infall region around the Virgo Cluster in
the spherical approximation (Karachentsev et al. 2014) is located with the
dotted circles.

\begin{figure*}

\clearpage

\includegraphics{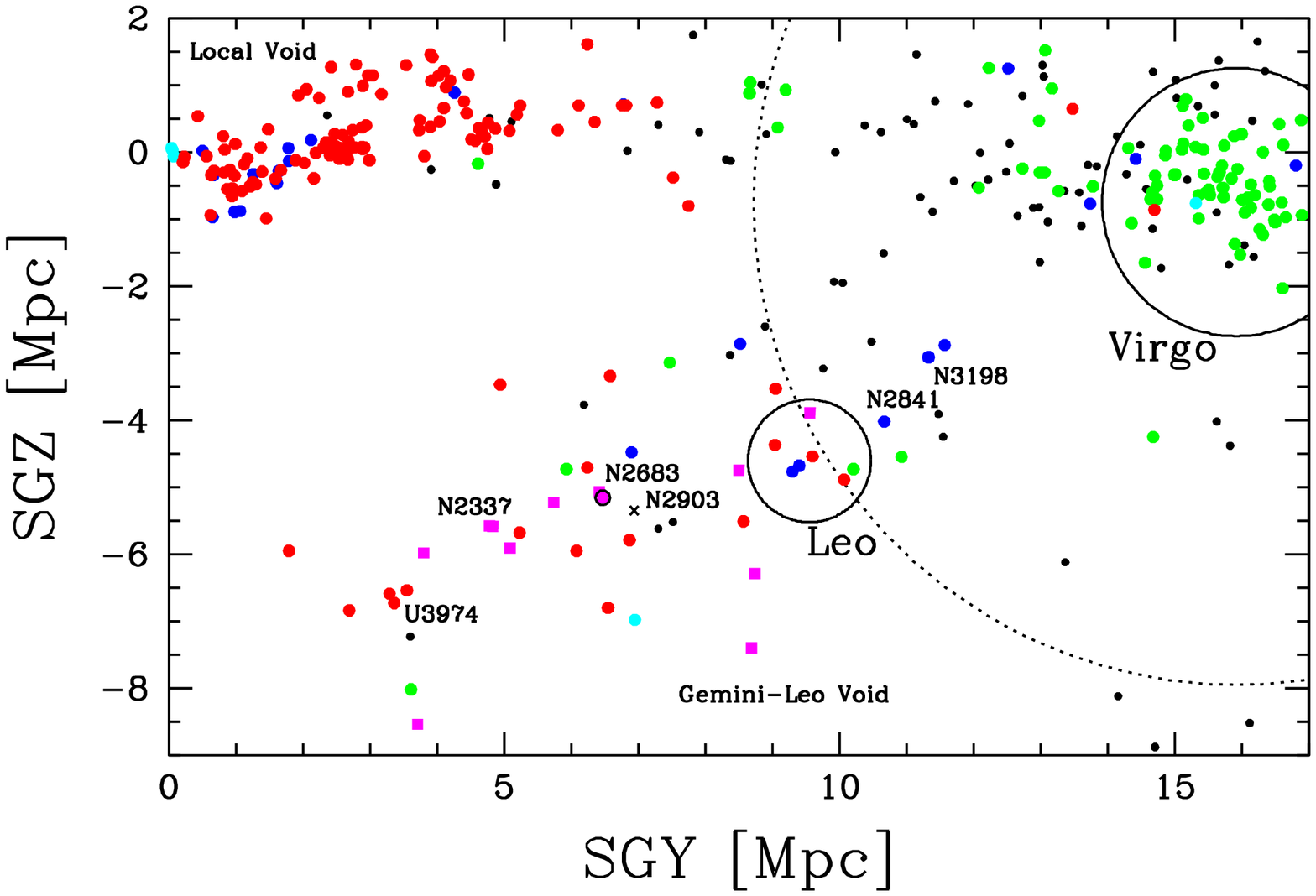}
\includegraphics{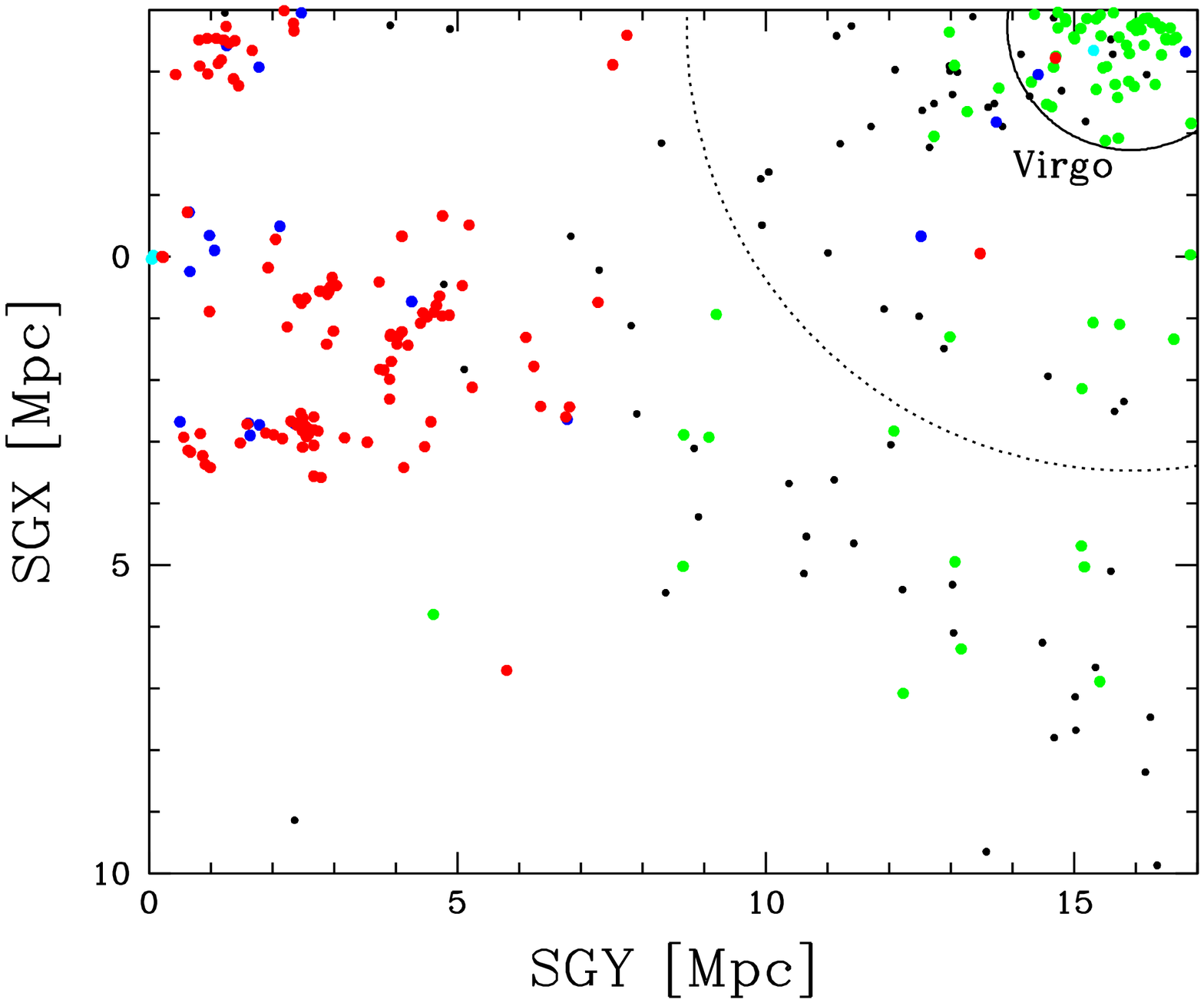}
\includegraphics{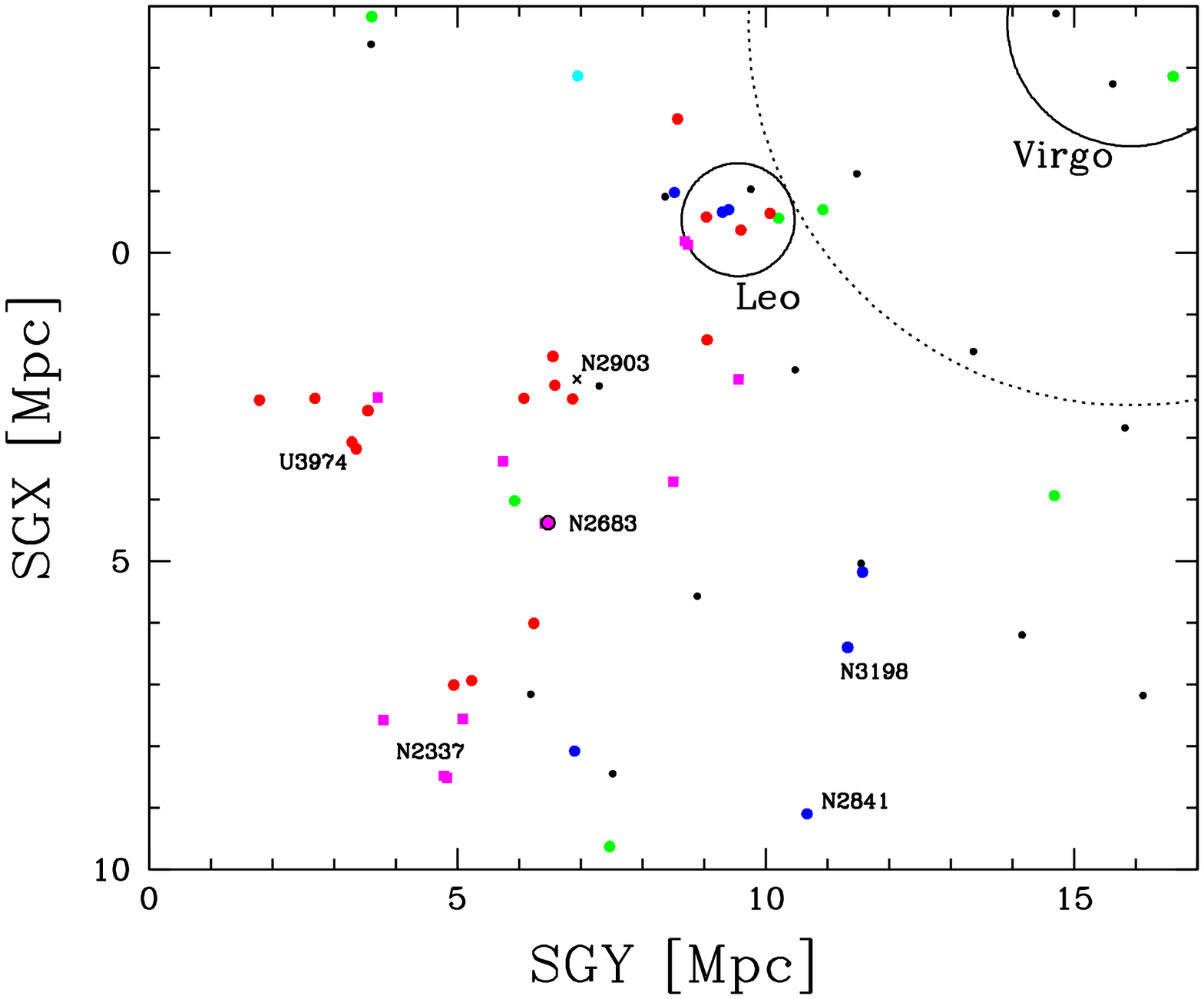}
\vspace{19cm}
\caption{}
\label{xyz}
\end{figure*}

Only galaxies with known distances are plotted because redshifts are not informative
due to severe departures from cosmic expansion.
Since nearer galaxies are more likely to have a known distance, the
impression of the distribution of galaxies is biased.  The representation
of the Leo Spur is skeletal.  
Nonetheless we see the important aspects.
The Leo Cluster is situated just outside the nominal edge
of the Virgo infall domain.   The region between the Leo and Virgo clusters
appears to be sparsely populated.
In the top panel of Figure \ref{xyz} we see that the Leo Spur 
lies in a band that is slightly sloped in SGZ roughly 5 Mpc below the Local Sheet.
There is a filament that runs slightly to the background from the modest
Leo Cluster to SGX$\sim 10$ Mpc at roughly constant SGY$\sim 11$ Mpc, 
and represented extremely sparsely here by three galaxies with Cepheid distances, 
NGC 2841, NGC 3198, and UGC 4284.  A second closer, hence better represented, strand runs 
through the galaxies NGC 2903\footnote{NGC 2903 does not yet have a good 
distance measurement; we give it the average distance of 2 neighbors.}  
and NGC 2683 to the region of the galaxy NGC 2337.  
This nearer domain is accessible to TRGB distance measurements with single
orbit observations with HST.  The Local Sheet and Leo Spur only partially
overlap in an SGX$-$SGY projection, with the Leo Spur running to larger values of SGX.
The nest of galaxies around UGC 3974 lies slightly apart from the nearer 
strand.  This group is of particular interest because of its location near the
supergalactic south pole where it receives almost the 
full reflex of our motion away from the Local Void.

The second and third sequences show the same scenes but add 
peculiar velocity vectors in two reference frames.  In that second sequence, 
Figure \ref{vpec},
the reference frame is the Local Sheet (TSK08) and cosmic expansion 
is removed assuming H$_0 = 74$ km/s/Mpc.  The reference frame is essentially
the same as the various versions of the Local Group frame.   Reasonable 
variations of the Hubble Constant have only small effects on the analysis.

The third sequence of Figure \ref{vres} is in what TSK08 called the Local Supercluster rest frame.
This reference frame nulled our motion with respect to all galaxies with measured 
distances within 3000 km/s.  The transformation from Local Sheet
coordinates can be separated into two vectors: a motion of 185 km/s directed 
toward the Virgo Cluster and a motion of 259 km/s directed almost due south
in supergalactic coordinates, away from the Local Void.

The two sequences demonstrating peculiar velocities reveal marked local 
coherence.   In the second sequence, that in a local frame of reference, it
is seen that galaxies within 8 Mpc lying within the supergalactic equatorial 
plane of the Local Sheet have only small
peculiar velocities.  Nearby galaxies in the Local Sheet are moving together 
in co-moving coordinates.  In this second sequence, essentially every other
galaxy, those beyond 8 Mpc or not in the Local Sheet, has a substantial peculiar 
velocity toward us.

In the third sequence, that in the Local Supercluster reference frame, the
pattern of motions is quite different.  In this reference frame, the Local Sheet 
is moving toward +SGY (toward Virgo) and $-$SGZ (away from the Local
Void).  Galaxies in the region of the Virgo Cluster have essentially zero velocity
in this reference frame. 
Along the Leo Spur the peculiar
velocities in the third sequence are mixed.  Motions are slightly positive 
at the end toward the Leo Cluster and increasingly negative at greater distances
from Virgo.  The positive velocities toward Virgo can be anticipated as due 
to the approach to the Virgo infall zone (not modeled here).  Negative peculiar 
velocities at the more distant regions from Virgo  could alternatively be due
to an underestimate of the motion of the Local Sheet away from the Local 
Void or to a peculiar motion of the Leo Spur toward positive SGZ because 
of a push from the Gemini$-$Leo Void.

The main conclusion of this section is that the Local Sheet motions identified 
in TSK08 are basically confirmed with, now, a substantial body of distance
measurements based on HST observations of the luminosities of tip of
the red giant branch sequences.  Inspection of the color-magnitude diagrams
that are presented will reveal that the tip locations are near the limit of the
methodology with one-orbit observations.  However there is no doubt that
the targets are more distant than expected from Hubble flow (tips would be 
brighter, hence more easily determined, if they were closer) and that peculiar
velocities in the Leo Spur, are strongly negative in the Local Sheet frame. 
With the data analyzed in this paper alone it can only be said that the 
Local Sheet and Leo Spur have motions toward each other in co-moving 
coordinates.  From consideration of data on larger scales it is clear that
the Local Sheet departs from the cosmic microwave background frame
with a motion away from the Local Void and toward the Leo Spur.  
Whether downward motion of the Local Sheet is the whole story or 
Leo Spur has an upward motion, as suggested by residuals  from the
TSK08 model seen in Figure~\ref{vres}, these possibilities await a new analysis of distance data over 
a much larger domain.

\section{Dwarf Associations}

The complementarity of all-sky searches for nearby galaxies (Karachentseva 
\& Karachentsev 1998 and following), accurate radial velocities from HI follow ups
(Huchtmeier et al. 2001), and HST TRGB distance measurements for
now almost 400 galaxies has led to a detailed picture of the structure
of groups in our vicinity.  Most galaxies in a volume limited sample are dwarfs.
Many dwarfs are found near giant galaxies but it is of considerable
interest that dwarfs also gather with other dwarfs.  There are very few 
extremely isolated dwarfs.

Tully et al. (2006) referred to regions crowded with dwarfs as associations 
and drew attention to seven such entities within 8 Mpc that share the following 
properties: no high luminosity 
member, $4-6$ dwarfs, scales of $\sim 300$ kpc, and radial velocity
dispersions $10-40$ km/s.  Characteristic crossing times are a substantial
fraction of the age of the universe so it is unlikely that the systems are 
in equilibrium, but if they are bound then their masses are in the range
$10^{11}-10^{12}~M_{\odot}$ and mass-to-blue-light values are several
hundred to a thousand.

The most distant of the associations identified in the 2006 paper 
lies in the Leo Spur
around the galaxy UGC 3974, labeled in Figs. \ref{xyz}$-$\ref{vres}.
Three tightly linked galaxies and an outrider were called the 14+19
Association\footnote{In Tully (1988b) the entity was erroneously 
linked to the 14 cloud, now called the Local Sheet, because of its
low systemic velocity; the Leo Spur is cloud 15 in that reference.}, 
their identification in the Nearby Galaxies Catalog (Tully 1988b). 
Today, three more dwarf galaxies have been identified 
in close proximity and with similar velocities.  Two of these, 
AGC 174585 and AGC 174605,
have TRGB distances.  The velocity dispersion of the 6 is very small,
21 km/s.   However the projected separations of the newly discovered 
companions are of order 700 kpc.   A halo with mass 
$5 \times 10^{11}~M_{\odot}$ is expected to have a virial domain 
of only 170/210 kpc (projected/3D) and an infall domain restricted to 500/610 kpc
(Tully 2015).
Only two galaxies in the ensemble, UGC 3974 and KK 65, lie in a common 
collapsed halo and only one more, UGC 4115, teeters on the edge of
infall toward those two.  UGC 3775 and the two AGC systems are each 
in separate halos and in expansion from the association unless there is
a very large amount of unseen matter.

The new distance information draws attention to an association around
NGC 2337, in an entity called 15+12 in the Nearby Galaxies Catalog.  There are four
galaxies in the region with TRGB distance measures\footnote{UGC 4426 
is a fifth galaxy but it is 2 Mpc from any of the others.} and two more
known with excellent velocity matches.  However the situation is 
similar to that of the 14+19 Association.  Only two galaxies, NGC 2337 and
UGC 3698, are likely to lie in a common halo and one more, UGC 3817,
probably lies within a related infall zone.  The velocity dispersion for these 
three is a minuscule 5 km/s in the line-of-sight.  Another pair, UGC 3860
and UGC 3966, probably lie within a common but distinct infall zone 
$\sim 1.5$ Mpc away.
UGC 3600 is expanding away from these others unless the 
dark matter concentration is in an extreme disproportion to the visible
matter.

\section{Concluding remarks}

The Leo Spur is the nearest distinct large scale structure outside of the
Local Sheet.  Systemic velocities in the Leo Spur are modest but already
with the identification of the local velocity anomaly (Faber \& Burstein 1988, Tully 1988a)
departures in this region from Hubble expansion were noticed.  Galaxies in the
Leo Spur are more distant than would be supposed from their velocities.
In the course of the analysis of the distance compilation {\it Cosmicflows-1} (TSK08),
a model was developed that provided an explanation.  A key observational
element is the {\it discontinuity} in peculiar velocities between those in the 
Local Sheet and those in the Leo Spur.  It was inferred that the Local Sheet
is descending toward the negative supergalactic pole due to the expansion of 
the Local Void.  The Leo Spur is decoupled from this expansion, hence, the
two structures have relative motions toward each other in co-moving coordinates.

While the distance information in TSK08 was limited, the observational situation
is considerably improved today with the publication of {\it Cosmicflows-2}
(Tully et al. 2013) and recent imaging with HST that results in the TRGB 
distance estimates presented here, supplement by contributions by 
McQuinn et al. (2014).  The new material emphatically confirms relevant
aspects of the TSK08 model.  The accumulated distance information 
demonstrates the spatial separation and kinematic discontinuity between
the Leo Spur and the Local Sheet.  The velocity field in the region embracing 
the Leo Spur, Local Sheet, and Virgo Cluster displays three overriding features
in the Local Sheet reference frame: (1) galaxies within the Local Sheet have 
very small peculiar velocities around the cosmic expansion, (2) galaxies in and
around the Virgo Cluster have large negative peculiar velocities, and 
(3) galaxies in the Leo Spur have large negative peculiar velocities.  
According to the TSK08 model, the galaxies of the Local Sheet are
responding together to two relatively local influences: an attraction toward the
Virgo Cluster causing a motion of 185 km/s and a repulsion away from the
Local Void causing a motion of 259 km/s.

\begin{figure*}
\includegraphics{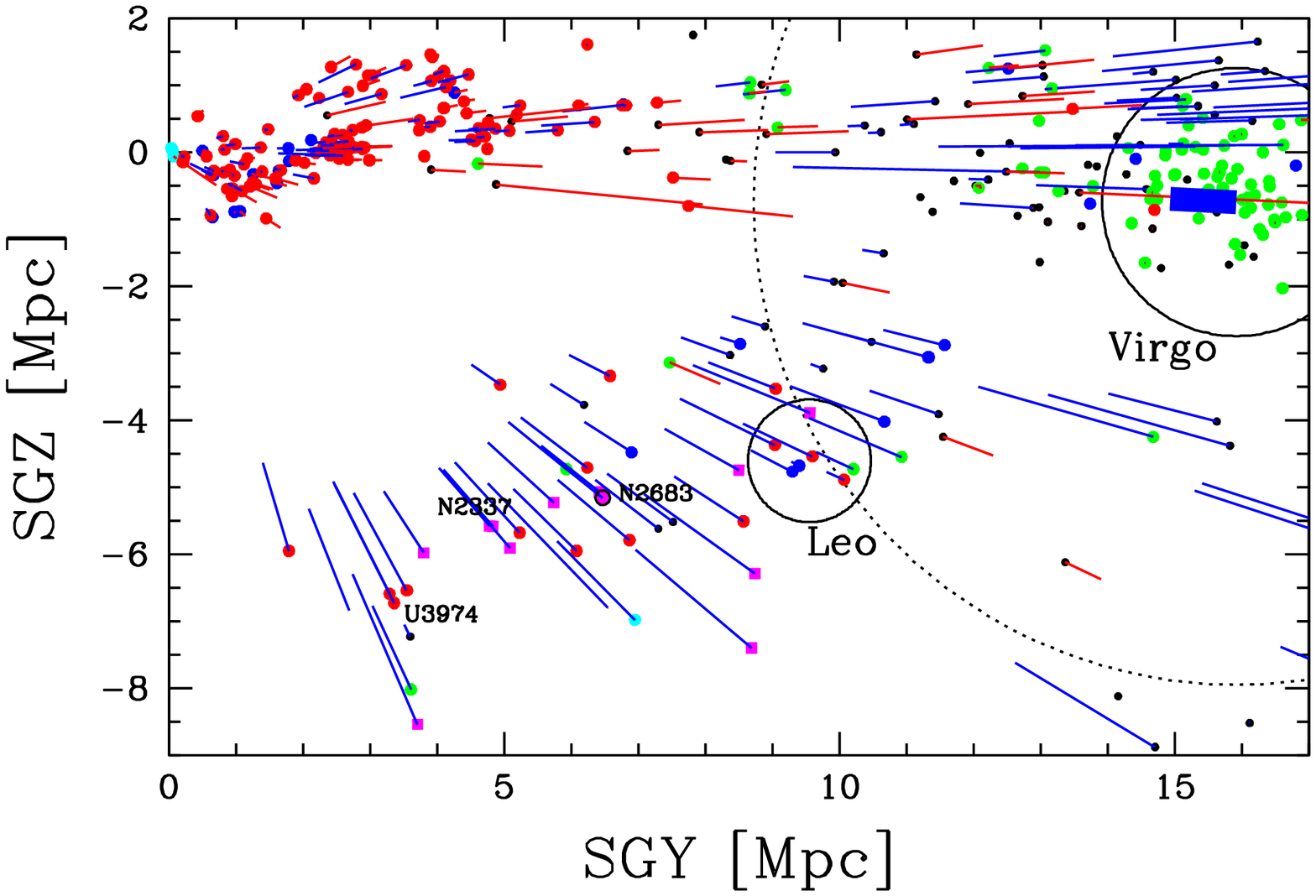}
\includegraphics{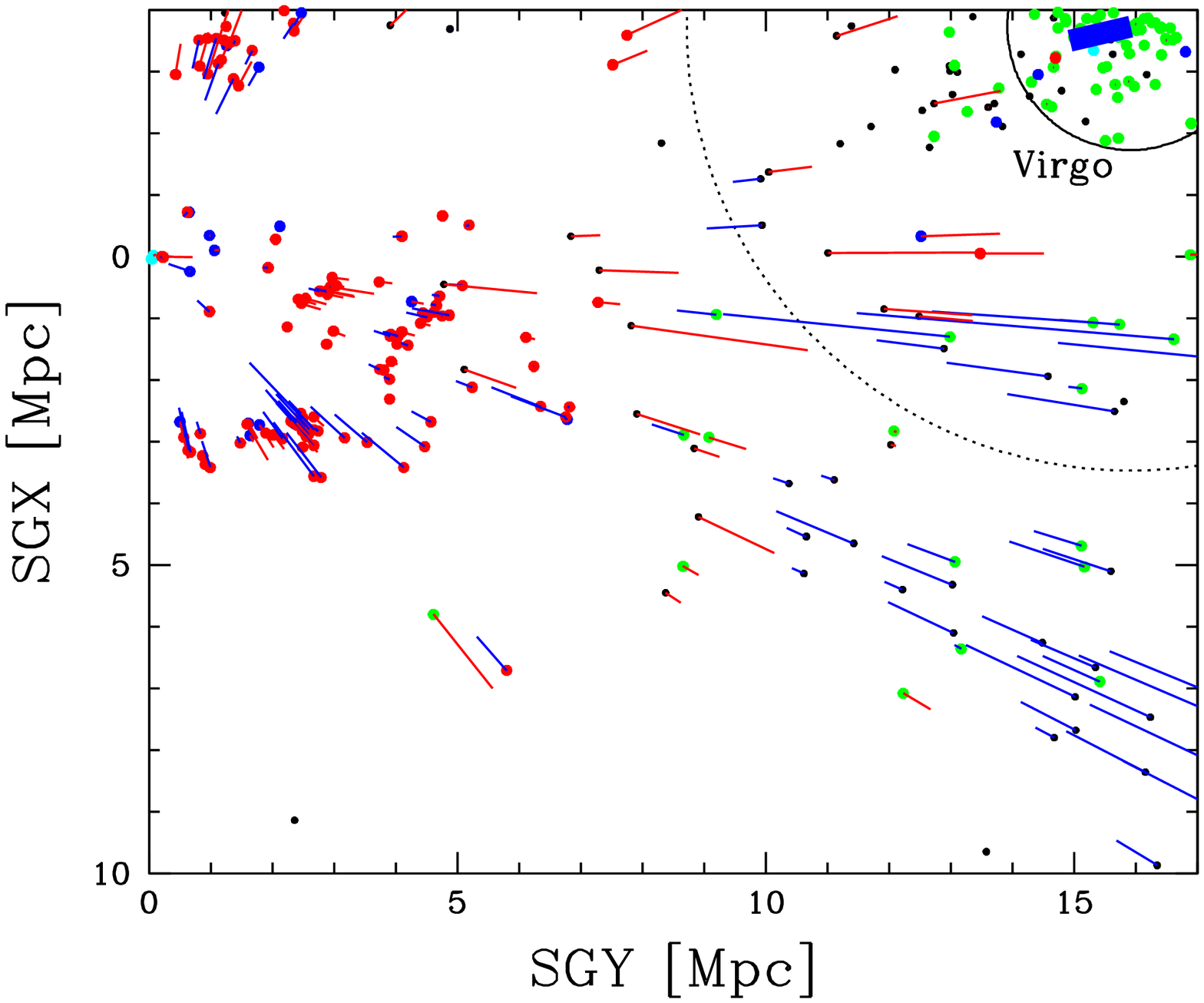}
\includegraphics{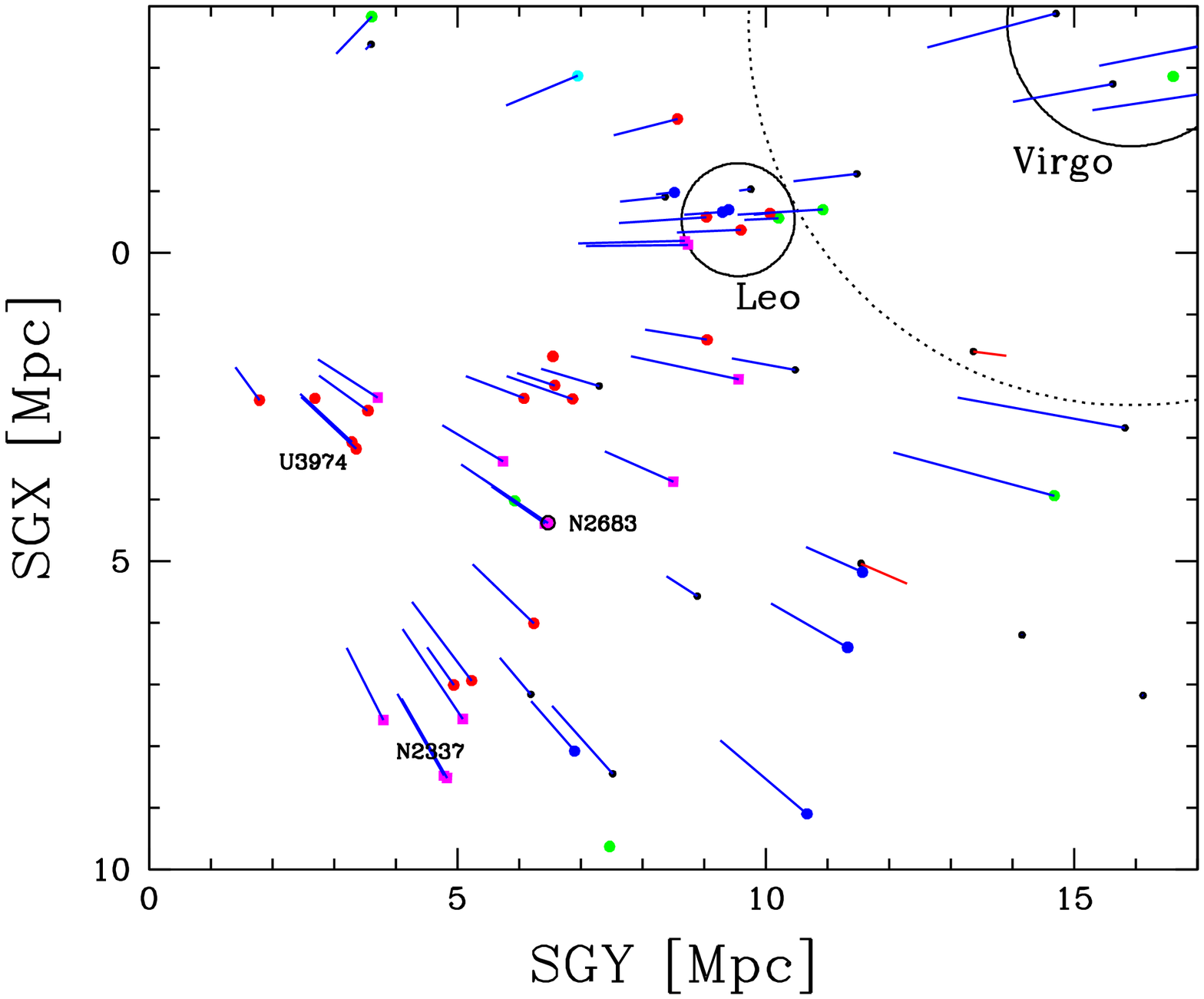}
\vspace{19cm}
\caption{}
\label{vpec}
\end{figure*}

\begin{figure*}
\includegraphics{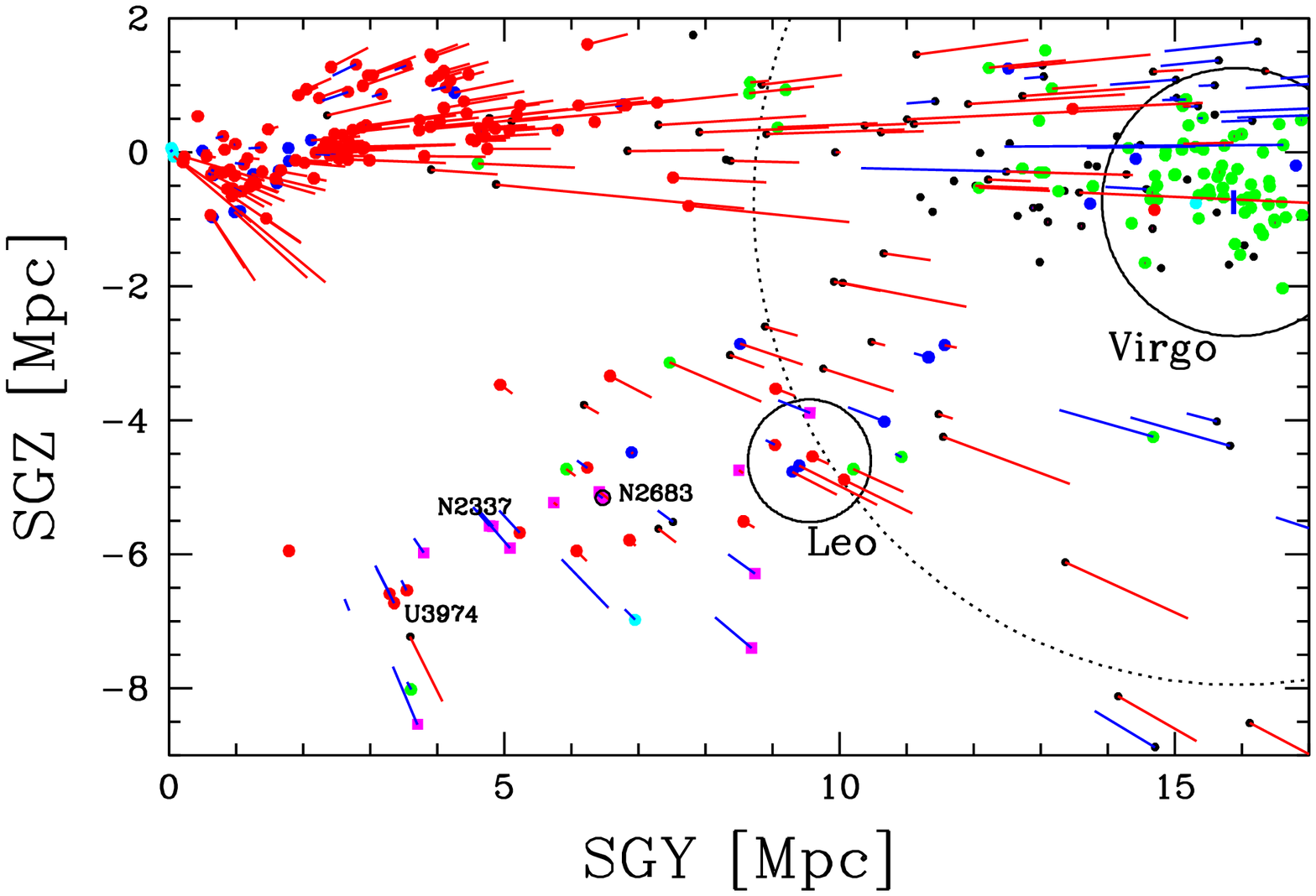}
\includegraphics{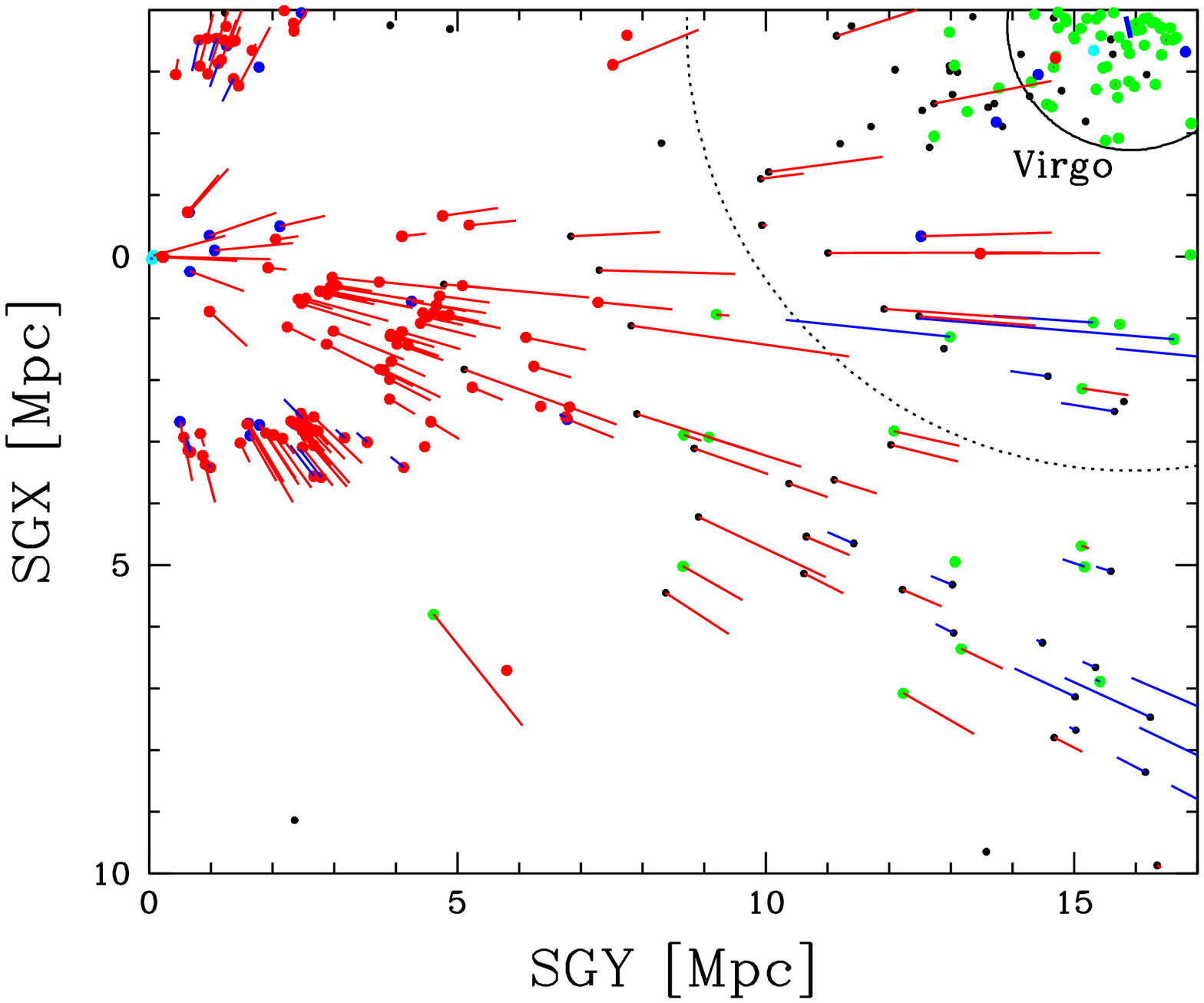}
\includegraphics{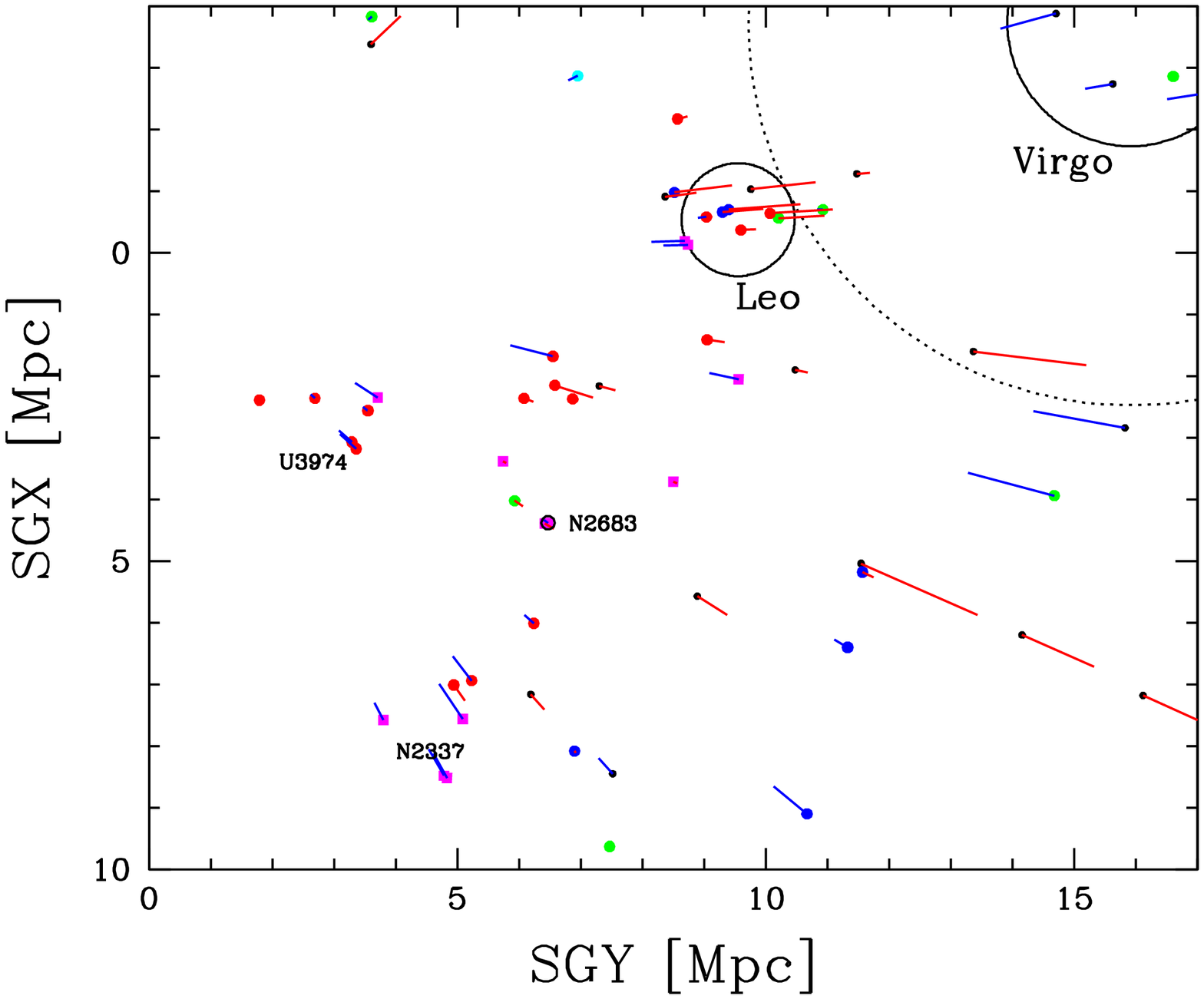}
\vspace{19cm}
\caption{}
\label{vres}
\end{figure*}

The reference frame referred to as `local supercluster' incorporates this
simple model.  It is seen in transitioning from the scenes of Fig. \ref{vpec}
to the scenes of Fig. \ref{vres} that (1) galaxies in the Local Sheet are
streaming toward the Virgo Cluster, (2) Virgo is practically at rest, and 
(3) the flow pattern in the Leo Spur has been
largely nulled out.  There is a slight trend along the Leo Spur toward positive
velocity residuals near the Leo Cluster and negative residuals at the far 
reaches from Leo.  Such a pattern can be explained by the attractive influence
of the Virgo Cluster near the Leo Cluster as the Virgo infall zone is approached
(Karachentsev et al. 2014) and the repulsive influence of the Gemini-Leo Void
that lies below the Leo Spur.  

In Tully et al. (2008) it was pointed out that a completely empty void in a topologically
flat $\Lambda$CDM universe with matter contributing 24\% of the critical density
expands at 16 km/s/Mpc.  This expansion rate was derived in two ways: analytically from 
the Friedman equation assuming spherical symmetry and from n-body simulations
(van de Weygaert \& Schaap 2007).  If a void is not empty, the expansion is reduced.  
It follows that the convergence of the Local Sheet and Leo Spur at $\sim 260$ km/s 
provides information regarding a combination of the size and emptiness of the voids 
at their edges.  This large motion implies large voids.  

It is to be appreciated that
the clear identification of peculiar motions away from voids is possible in the present
situation because (1) individual peculiar velocities are large compared with uncertainties
coming from distance errors and (2) the geometry is such that we can see the full
amplitude of motions in radial velocities.  We are afforded insight into the properties of voids 
due to our proximity to the Local Void that is otherwise not easily available.  

The 2008 model deserves to be refined in the light of the new distance information.
However, if only peculiar velocities within the restricted domain explored in this 
paper are considered then only the relative convergence of the Local Sheet, 
Leo Spur and Virgo Cluster components can be evaluated; but not the relative
importance of the external influences.   To properly evaluate the relative influences 
of the downward push of the Local Void versus the upward push of the Gemini-Leo Void, 
it will be necessary to 
give attention to distances and the velocity field over a much larger volume, 
a task for later study.

As the name implies, the Leo Spur is not a major feature.  The strand nearest us
that is most easily studied peters out into associations of dwarfs.  
Such regions are of interest because they may be the visible manifestations 
of halos at masses below $10^{12}~M_{\odot}$.  Dwarf galaxies that lie apart from 
major galaxies tend to flock together.  However, in the two cases presented in the 
Leo Spur, the 14+19 and 15+12 associations, only a few of the dwarfs are likely 
to be bound to each other unless there is a lot more dark matter in their vicinity 
than supposed. 

\bigskip\noindent	       
{\bf Acknowledgements:}
This work is based
on observations made with the NASA/ESA Hubble Space
Telescope. STScI is operated by the Association of Universities
 for Research in Astronomy, Inc. under NASA contract
NAS 526555. The work in Russia is supported by RFBR
grants 13--02--90407 and 13--02--92960. L.N.M. and D.I.M. acknowledge 
support from RFBR grant 13--02--00780 and Research Program OFN17 of the 
Division of Physics, Russian Academy of Sciences.

\clearpage
\noindent
{\bf References} 

	

\noindent	
Brinckmann T., Lindholmer M., Hansen S.H., Falco M., arXiv:1411.6650


\noindent	
Cautun M., van de Weygaert R., Jones B.J.T., Frenk C.S., 2014, MNRAS, 441, 2923

\noindent	
Dolphin, A. 2000, PASP, 112, 1383

\noindent	
Elyiv A.A., Karachentsev I.D.,Karachentseva V.E., Melnyk, O.V., Makarov D.I., 2013, Astophys. Bulletin, 68, 1 

\noindent	
Faber S.M., Burstein D., 1988, in Large-Scale Motions in the Universe (Princeton: Princeton Univ. Press), 115

\noindent	
Huchtmeier, W.K., Karachentsev, I.D., Karachentseva, V.E. 2001, A\&A, 377, 801

\noindent	
Jacobs, B.A., Rizzi, L, Tully, R.B. et al. 2009, AJ, 138, 332

\noindent	
Jacobs, B.A., Tully, R.B., Rizzi, L., et al. 2011, AJ, 141, 106


\noindent
Karachentsev, I.D., Kaisina, E.I. 2013, AJ, 146, 46


\noindent	
Karachentsev I.D., Makarov D.I., Kaisina E.I., 2013, AJ, 145, 101 (UNGC)

\noindent	
Karachentsev I.D., Makarov D.I., Sharina M.E., et al., 2003, A \& A, 398, 479

\noindent	
Karachentsev, I.D., Tully, R.B., Wu, P.-F., et al. 2014, ApJ, 782, 4

\noindent	
Karachentseva, V.E., Karachentsev, I.D. 1998, A\&AS, 127, 409



\noindent	
Makarov, D.I, Makarova, L., Rizzi, L. et al. 2006, AJ, 132, 2729

\noindent	
McQuinn K.B.W., Cannon J.M., Dolphin A.E. et al. 2014, ApJ, 785, 3

\noindent	
Peebles P.J.E., 1980, The Large-scale structure of the universe, Princeton University Press 



\noindent	
Rizzi L., Tully R.B., Makarov D.I., et al., 2007, ApJ, 661, 813

\noindent	
Schlafly, E.F., Finkbeiner, D.P., 2011, ApJ, 737, 103

\noindent	
Shandarin S., Habib S., Heitmann K., 2012, Physical Review D., 85, 8

\noindent	
Shandarin S. F., Klypin A. A., 1984, SvA, 28, 491

\noindent	
Shandarin S. F., Zeldovich Y. B., 1989, Rev. Mod. Phys., 61, 185


\noindent	
Tully R.B., 2015, AJ, 149, 54

\noindent	
Tully R.B., Courtois H.M., Dolphin A.E. et al. 2013, AJ, 146, 86 

\noindent	
Tully, R.B., Fisher, J.R. 1987, Nearby Galaxies Atlas (Cambridge Univ. Press)

\noindent	
Tully R.B., Shaya E.J., Karachentsev I.D., Courtois H.M., Kocevski D.D.,
Rizzi, L., \& Peel, A. 2008, ApJ, 676, 184 (TSK08)
	    
\noindent	
Tully R.B., Rizzi L., Dolphin A.E., Karachentsev I.D. et al., 2006, AJ, 132, 729

\noindent	
Tully R.B., 1988a, in Large-Scale Motions in the Universe (Princeton: Princeton Univ. Press), 169

\noindent	
Tully R.B., 1988b, Nearby Galaxies Catalog (Cambridge Univ. Press)

\noindent	
Zeldovich, Ya.B., 1970, A\& A 5, 84

\noindent
van de Weygaert, R., Schaap, W., 2007, in Data Analysis in Cosmology, ed. V. Martinez et al. (Berlin: Springer), arXiv:0708.1441

\noindent	
van Haarlem, M., van de Weygaert, R., 1993, ApJ, 418, 544

     \clearpage

\clearpage

 \noindent
 Fig. 3 caption. Color-magnitude diagrams for target galaxies from ACS
 observations. The dotted horizontal lines mark the magnitudes of
 the TRGB.

\clearpage

 \noindent
 Fig. 4 caption. Projections of the Leo Spur region in supergalactic coordinates.
 Top: SGY$-$SGZ view with $-4 < SGX < +10$ Mpc. 
 Mid and bottom: SGX$-$SGY views split $+2>SGZ>-2$ and
 $-2>SGZ>-9$ Mpc to
 isolate the Local Sheet in the middle panel and the Leo Spur
 in the bottom panel.
 Symbol colors identify sources of distance measurements:
 blue $-$ cepheid; green $-$ SBF; red $-$ previous TRGB; 
 magenta squares $-$ new TRGB.
 Important galaxies are identified.  
 Solid circles: Virgo and Leo cluster virial regions.
 Dotted circles: Virgo Cluster infall region.
 The Local and Gemini-Leo voids are extensive regions above the Local Sheet and below 
 the Leo Spur, respectively, in SGZ.
 
\clearpage

\clearpage

 \noindent
 Fig. 5 caption. Same as Fig. 4 with the addition of line-of-sight peculiar velocity vectors
 in the Local Sheet reference frame, assuming H$_0=74$ km/s/Mpc.  
 Negative peculiar velocity vectors are blue and toward us
 while positive peculiar velocity vectors are red and directed away.
 The heavy blue vector at the Virgo Cluster is an average for the entire cluster. 
 Scale: vector 1 Mpc in length = 200 km/s.
\clearpage

\clearpage

 \noindent
 Fig. 6 caption. Same as Fig. 5 except
 in the Local Supercluster reference frame, assuming H$_0=74$ km/s/Mpc.  
 Negative peculiar velocity vectors are blue and toward us
 while positive peculiar velocity vectors are red and directed away.
 The averaged Virgo Cluster vector is only slightly negative. 
 Scale: vector 1 Mpc in length = 200 km/s.
\clearpage

\begin{table}
{\scriptsize
\caption{Galaxies in the Leo Spur recently observed with HST.}
\begin{tabular}{lrccccccrcrccr} 
\hline

 Name    & PGC  & RA (J2000) Dec   &   SGL    &  SGB  &$V_h$&$V_{LS}$&$V_{LSC}$&$a_{Ho}$&$B_T$&$T$&$I_{TRGB}$&$A_I$&$D_{TRGB}$\\
\hline
UGC3600   &  19871 &065540.0+390542&  26.60 & -35.20 & 412 & 448 & 693 & 1.91 &16.2  & 8 & 26.17 & 0.17 & 10.38$\pm$1.02\\
UGC3698   &  20264 &070918.8+442248&  29.39 & -29.82 & 422 & 479 & 716 & 1.17 &15.4  &10 & 26.36 & 0.18 & 11.22$\pm$1.11\\
NGC2337   &  20298 &071013.6+442725&  29.58 & -29.73 & 436 & 493 & 730 & 2.75 &13.5  & 9 & 26.36 & 0.17 & 11.28$\pm$0.57\\
UGC3860   &  21073 &072817.2+404613&  33.96 & -33.00 & 354 & 386 & 641 & 1.41 &15.1  &10 & 26.18 & 0.11 & 10.86$\pm$1.34\\
NGC2683   &  24930 &085240.9+332502&  55.87 & -33.42 & 411 & 381 & 673 &13.49 &10.3  & 3 & 25.91 & 0.06 &  9.36$\pm$0.28\\
KK69      & 166095 &085250.7+334752&  55.64 & -33.09 & 463 & 435 & 726 & 1.38 &17.4  &10 & 25.82 & 0.06 &  9.28$\pm$0.28\\
KK70      & 166096 &085522.0+333333&  56.32 & -32.98 &  -- &  -- &  ---& 0.50 &17.7  &-3 & 25.79 & 0.06 &  9.18$\pm$0.30\\
UGC5209   &  27935 &094504.2+321418&  66.40 & -27.11 & 538 & 498 & 787 & 0.83 &16.1  &10 & 26.03 & 0.04 & 10.42$\pm$0.35\\
UGC5288   &  28378 &095117.2+074938&  91.28 & -40.42 & 556 & 390 & 703 & 1.45 &14.6  & 9 & 26.30 & 0.07 & 11.41$\pm$1.10\\
UGC5456   &  29428 &100719.6+102143&  90.82 & -35.73 & 544 & 391 & 699 & 1.62 &13.8  & 9 & 26.08 & 0.06 & 10.77$\pm$1.07\\
AGC174605 &5060076 &075021.7+074740&  57.66 & -62.79 & 351 & 206 & 517 & 0.34 &18.0  &10 & 25.83 & 0.04 &  9.60$\pm$0.18\\
AGC182595 &4087020 &085112.1+275248&  59.52 & -38.12 & 396 & 336 & 640 & 0.36 &17.2  & 9 & 25.61 & 0.07 &  8.47$\pm$0.14\\
AGC731457 &1824266 &103155.8+280134&  77.87 & -21.68 & 454 & 396 & 675 & 0.38 &16.8  &10 & 26.07 & 0.04 & 10.52$\pm$0.34\\
\hline
\end{tabular}
}
\end{table}

\end{document}